\documentclass[reprint,amssymb,amsmath,superscriptaddress,aps,showpacs,10pt,floatfix,longbibliography,prx]{revtex4-1}

\usepackage[pdftex]{graphicx} 
\usepackage{epstopdf}
\usepackage{verbatim}
\usepackage{amsmath}
\usepackage{color}
\usepackage{subfigure}
\usepackage{amsbsy}
\usepackage{wasysym}

\begin{document}

\title{Partial up-up-down order with the continuously distributed order parameter in the triangular antiferromagnet TmMgGaO$_4$}

\author{Yuesheng Li}
\email{yuesheng\_li@hust.edu.cn}
\affiliation{Experimental Physics VI, Center for Electronic Correlations and Magnetism, University of Augsburg, 86159 Augsburg, Germany}
\affiliation{Wuhan National High Magnetic Field Center and School of Physics, Huazhong University of Science and Technology, 430074 Wuhan, China}

\author{Sebastian Bachus}
\affiliation{Experimental Physics VI, Center for Electronic Correlations and Magnetism, University of Augsburg, 86159 Augsburg, Germany}

\author{Hao Deng}
\email{hao.deng@frm2.tum.de}
\affiliation{Institute of Crystallography, RWTH Aachen University and J\"{u}lich Centre for Neutron Science (JCNS) at Heinz Maier-Leibnitz Zentrum (MLZ), 85748 Garching, Germany}

\author{Wolfgang Schmidt}
\affiliation{Forschungszentrum J\"{u}lich GmbH, J\"{u}lich Centre for Neutron Science at ILL, 71 Avenue des Martyrs, 38042 Grenoble, France}

\author{Henrik Thoma}
\affiliation{Institute of Crystallography, RWTH Aachen University and J\"{u}lich Centre for Neutron Science (JCNS) at Heinz Maier-Leibnitz Zentrum (MLZ), 85748 Garching, Germany}

\author{Vladimir Hutanu}
\affiliation{Institute of Crystallography, RWTH Aachen University and J\"{u}lich Centre for Neutron Science (JCNS) at Heinz Maier-Leibnitz Zentrum (MLZ), 85748 Garching, Germany}

\author{Yoshifumi Tokiwa}
\affiliation{Experimental Physics VI, Center for Electronic Correlations and Magnetism, University of Augsburg, 86159 Augsburg, Germany}

\author{Alexander A. Tsirlin}
\affiliation{Experimental Physics VI, Center for Electronic Correlations and Magnetism, University of Augsburg, 86159 Augsburg, Germany}

\author{Philipp Gegenwart}
\email{philipp.gegenwart@physik.uni-augsburg.de}
\affiliation{Experimental Physics VI, Center for Electronic Correlations and Magnetism, University of Augsburg, 86159 Augsburg, Germany}

\date{\today}

\begin{abstract}
Frustrated quasidoublets without time-reversal symmetry can host highly unconventional magnetic structures with continuously distributed order parameters even in a single-phase crystal. Here, we report the comprehensive thermodynamic and neutron diffraction investigation on the single crystal of TmMgGaO$_4$, which entails non-Kramers Tm$^{3+}$ ions arranged on a geometrically perfect triangular lattice. The crystal electric field (CEF) randomness caused by the site-mixing disorder of the nonmagnetic Mg$^{2+}$ and Ga$^{3+}$ ions, merges two lowest-lying CEF singlets of Tm$^{3+}$ into a ground-state (GS) quasidoublet. Well below $T_c$ $\sim$ 0.7 K, a small fraction of the antiferromagnetically coupled Tm$^{3+}$ Ising quasidoublets with small inner gaps condense into two-dimensional (2D) up-up-down magnetic structures with continuously distributed order parameters, and give rise to the \emph{columnar} magnetic neutron reflections below $\mu_0H_c$ $\sim$ 2.6 T, with highly anisotropic correlation lengths, $\xi_{ab}$ $\geq$ 250$a$ in the triangular plane and $\xi_c$ $<$ $c$/12 between the planes. The remaining fraction of the Tm$^{3+}$ ions remain nonmagnetic at 0 T and become uniformly polarized by the applied longitudinal field at low temperatures. We argue that the similar model can be generally applied to other compounds of non-Kramers rare-earth ions with correlated GS quasidoublets.
\end{abstract}

\maketitle

\section{Introduction}

Geometrical frustration can render the ground state(s) of the correlated spin system macroscopically degenerate and completely disordered in the classical Ising case~\cite{wannier1950antiferromagnetism,kano1953antiferromagnetism,bradley2019robust,ramirez1999zero,bramwell2001spin,morris2009dirac}, or trigger strong quantum fluctuations that prevent the conventional symmetry breaking even down to $\sim$ 0 K in the quantum case~\cite{anderson1973resonating,moessner2006geometrical,balents2010spin,li2015gapless,li2015rare,shores2005structurally,shimizu2003spin,itou2008quantum,li2014gapless}. In most of the previously proposed frustrated magnets, the (effective) $S$ = 1/2 dipole moments of the ground-state (GS) doublets are protected either by time-reversal symmetry in the case of Kramers ions with an odd number of electrons per site~\cite{ramirez1999zero,PhysRevLett.118.107202,shores2005structurally,shimizu2003spin,itou2008quantum,li2014gapless} or by the local crystal electric field (CEF) symmetry in the case of non-Kramers ions with an even number of electrons per site~\cite{PhysRevB.91.224430,PhysRevB.94.024436,PhysRevLett.105.047201,PhysRevB.83.094411}. In non-Kramers ions without symmetry-protected doublets, two close-lying singlets will typically occur~\cite{PhysRev.172.539}, as in the three-dimensional dipolar Ising ferromagnet LiTbF$_4$~\cite{PhysRevB.12.180,PhysRevLett.37.1161,PhysRevB.12.191} and in the kagome magnet Pr$_3$Ga$_5$SiO$_{14}$~\cite{PhysRevB.81.224416}. However, geometrically frustrated non-Kramers magnets with correlated GS quasidoublets are still rare to date. Here, the complete site-mixing disorder between two nonmagnetic ions with different valences significantly distributes the energies of the two lowest-lying and nearly degenerate CEF singlets. And the intersite spin interactions combined with the single-ion terms and the randomness can lead to exotic quantum phases at low temperatures~\cite{moessner2000two}.

In a search for such a kind of material with the two-dimensional (2D) triangular arrangement of the $4f$ ions, we explored structural siblings of YbMgGaO$_4$, which we recently characterized as a quantum spin liquid (QSL) candidate with the experimental evidence for breaking and re-arrangement of uncorrelated/resonating valence bonds~\cite{PhysRevLett.117.097201,li2017nearest,PhysRevLett.122.137201}. Despite the disorder effect on the CEF and the putative QSL state~\cite{PhysRevLett.122.137201,zhu2017disorder,kimchi2017valence} caused by the site-mixing disorder between nonmagnetic Mg$^{2+}$ and Ga$^{3+}$ ions in YbMgGaO$_4$, the GS CEF doublet of the Kramers Yb$^{3+}$ ion always gives rise to the effective $S$ = 1/2 magnetic moment at $T$ $\ll$ $\Delta_{CEF}$/$k_B$ $\sim$ 460 K, where $\Delta_{CEF}$ is the energy gap to the first excited level~\cite{PhysRevLett.118.107202}. While, the many-body correlated physics may be significantly changed when Yb$^{3+}$ is replaced by the non-Kramers Tm$^{3+}$ (4$f^{12}$) ion on the triangular lattice, as the time reversal symmetry is no longer preserved, and the previously symmetry-protected degeneracy of the GS CEF doublet may get ``lifted". And thus exotic quantum phases may emerge in the new frustrated magnet, TmMgGaO$_4$. Recently, single crystals of TmMgGaO$_4$ were successfully synthesized by Cevallos \emph{et al.}~\cite{cevallos2017anisotropic}, which provides an opportunity to study its exotic correlated magnetism experimentally.

In this paper, we report a thorough single-crystal investigation of the low-temperature magnetism of TmMgGaO$_4$, including heat capacity, Faraday force magnetization (susceptibility), magnetocaloric effect, and neutron diffraction measurements, down to 30 mK. The Mg$^{2+}$/Ga$^{3+}$ disorder significantly distributes the energies of the two lowest-lying CEF singlets, thus mixing them into a GS quasidoublet. At low temperatures and in small longitudinal fields, a fraction of the Tm$^{3+}$ ions -- those characterized by a small gap between the two CEF singlets -- give rise to the novel 2D Ising up-up-down (uud) phase with the continuously distributed order parameter, under the frustrated intersite couplings on the triangular lattice. The remaining large fraction with the large inner gaps remains nonmagnetic at 0 T and becomes uniformly polarized by the applied longitudinal field. Using the random many-body-correlated model of the GS quasidoublets, we can naturally interpret most of the low-$T$ magnetic properties. A similar model can be generally applied to other non-Kramers rare-earth magnets with correlated GS quasidoublets.

\section{Technical details}

High-quality single crystals ($\sim$ 1 cm) of TmMgGaO$_4$, Tm$_{0.04}$Lu$_{0.96}$MgGaO$_4$, and Yb$_{0.04}$Lu$_{0.96}$MgGaO$_4$ were grown by the
floating zone technique (Appendix~\ref{a1})~\cite{li2015rare,cevallos2017anisotropic}. The Faraday force magnetization~\cite{sakakibara1994faraday}, heat capacity, magnetocaloric effect (magnetic Gr\"{u}neisen ratio) ~\cite{tokiwa2011high,tokiwa2014quantum} down to 30 mK were measured in a $^3$He-$^4$He dilution refrigerator (Appendix~\ref{a2}). The neutron diffraction experiments were carried out in the $ab$ plane ($L$ = 0) and along the $c$ axis ($L$ $\neq$ 0), on the CEA-CRG single crystal diffractometer D23~\cite{ressouche1999new} of Institut Laue-Langevin (ILL) in France and on the single crystal diffractometer POLI~\cite{hutanu2015poli} of Heinz Maier-Leibnitz Zentrum (MLZ) in Germany, respectively, down to 60 mK and up to 5 T. Using the Matlab codes, we performed CEF, exact diagonalization (ED), and spin-wave calculations for a model spin Hamiltonian. Then we simultaneously fit this model to the temperature dependence of direct-current (dc) susceptibility and heat capacity measured at $\sim$ 0 T, as well as the field dependence of magnetization at 40 mK, by minimizing the following function,
\begin{equation}
R_p=\sqrt{\frac{1}{N_0}\sum_{i}(\frac{X_i^{obs}-X_i^{cal}}{\sigma_i^{obs}})^2},
\label{Eq1}
\end{equation}
where $N_0$, $X_i^{obs}$ and $\sigma_i^{obs}$ are the number of the data points, the observed value and its standard deviation, respectively, whereas $X_i^{cal}$ is the calculated value.

\begin{figure*}[t]
\begin{center}
\includegraphics[width=18cm,angle=0]{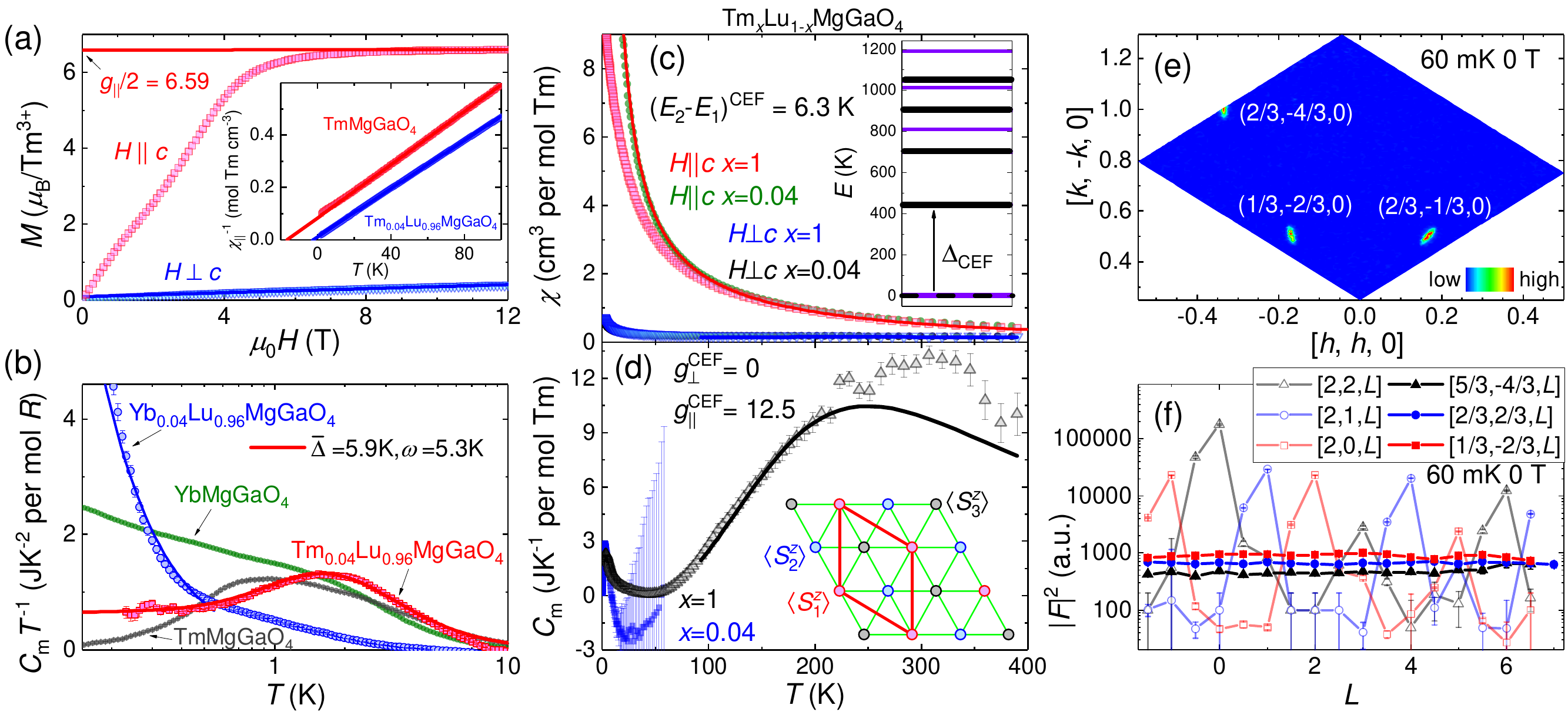}
\caption{(Color online)
(a) Magnetization of TmMgGaO$_4$ measured at 1.9 K in the fields parallel and perpendicular to the $c$ axis. The colored lines show the linear fits to the data above 8 T. Inset: Curie-Weiss fits to the susceptibilities of TmMgGaO$_4$ and Tm$_{0.04}$Lu$_{0.96}$MgGaO$_4$ measured at 0.1 T along the $c$ axis. (b) Magnetic heat capacities ($C_m$) of TmMgGaO$_4$, Tm$_{0.04}$Lu$_{0.96}$MgGaO$_4$, YbMgGaO$_4$, and Yb$_{0.04}$Lu$_{0.96}$MgGaO$_4$ at 0 T. The red and blue lines show, respectively, the fits to the data for Tm$_{0.04}$Lu$_{0.96}$MgGaO$_4$ and Yb$_{0.04}$Lu$_{0.96}$MgGaO$_4$ with the Lorentzian distributions of $E_2$-$E_1$. (c) Temperature dependence of susceptibilities measured in the field of 0.05 T applied both parallel and perpendicular to the $c$ axis. The inset shows the CEF levels from the combined CEF fit, with the black and violet lines for the CEF doublets and singlets, respectively. The GS quasidoublet is shown by the black-violet line. (d) Temperature dependence of the magnetic heat capacity measured at 0 T. The inset presents sketch of the 2D three-sublattice magnetic dipole structure with the green and red lines showing the triangular lattice and the magnetic unit cell, respectively. The lines show the combined CEF fit to the magnetic data of TmMgGaO$_4$ above 90 K in both (c) and (d). (e) Magnetic neutron diffraction of TmMgGaO$_4$ measured on D23 at 60 mK and 0 T in the $ab$ plane ($L$ = 0). (f) $L$ dependence of selected static structure factors measured on POLI at 60 mK and 0 T. The magnetic structure factors are normalized by the magnetic form factor of Tm$^{3+}$.}
\label{fig1}
\end{center}
\end{figure*}

\section{Single-ion physics}

Generally, the CEF of Tm$^{3+}$ with the $D_{3d}$ point-group symmetry of TmMgGaO$_4$ splits the 13-degenerate GS of the free Tm$^{3+}$ ion with the total angular momentum $J$ = 6, $\mid$$m_J$$\rangle$ ($m_J$ = 0, $\pm$1..., $\pm$$J$), into five singlets (3$A_{1g}$+2$A_{2g}$) and four doublets (4$E_g$), according to the symmetry analysis. In the following, we will further determine the low-lying CEF states of Tm$^{3+}$ in TmMgGaO$_4$ by thermodynamical measurements.

At $\sim$ 1.9 K, the effective spin-1/2 moments of Tm$^{3+}$ can be fully polarized above $\sim$ 8 T applied along the $c$ axis [see Fig.~\ref{fig1} (a)]. Through a linear fit to this high-field magnetization data, we obtain the fitted intercept that measures the saturated effective spin-1/2 magnetic moment $g_{\parallel}$/2, where $g_{\parallel}$ = 13.18(1), and the small slope that corresponds to the van Vleck susceptibility, $\chi_{\parallel}^{vv}$ = 0.003(1) cm$^3$/mol~\cite{li2015gapless,li2015rare}. While, along the $ab$ plane the magnetization shows a linear field dependence between 0 and 12 T with the nearly zero intercept [see Fig.~\ref{fig1} (a)], suggesting the strict Ising anisotropy of the Tm$^{3+}$ magnetic moments (Appendix~\ref{a3}). Between $\sim$ 30 and 60 K, the magnetic entropy of TmMgGaO$_4$ is measured to be constant, $S_m$ $\sim$ $R$ln2 per mole [see Fig.~\ref{fig2} (e)], confirming the GS CEF (quasi)doublets, and thus the formation of the effective Ising spin-1/2 moments of Tm$^{3+}$ below 60 K (Appendix~\ref{a3}).

This Ising nature is rooted in the GS CEF quasidoublets~\cite{nekvasil1990effective}. To better understand its nature, we first prepared the highly diluted samples of Tm$_x$Lu$_{1-x}$MgGaO$_4$ ($x$ = 0.04), where no intersite interactions occur, and single-ion physics of Tm$^{3+}$ can be probed. The diluted Yb$_x$Lu$_{1-x}$MgGaO$_4$ sample with the Yb$^{3+}$ Kramers ion was also studied as reference. In both cases, the dilution eliminates any intersite magnetic couplings, as confirmed by the diminutively small Curie-Weiss temperatures, $\theta_w^{\parallel}$($x$ = 0.04) $\sim$ 0.16$\theta_w^{\parallel}$($x$ = 1) [see Fig.~\ref{fig1} (a) for Tm$_x$Lu$_{1-x}$MgGaO$_4$ and Ref.~\cite{li2015gapless} for Yb$_x$Lu$_{1-x}$MgGaO$_4$] . The small Curie-Weiss temperature is obtained by the fit to $\chi_{\parallel}$ measured in the temperature range where $S_m$ $\sim$ $R$ln2, and thus the CEF effect due to excitations to higher CEF levels is negligible, and we obtain $\theta_w^{\parallel}$ = -3$\overline{(J_1^{zz}+J_2^{zz})}$/2 (see below)~\cite{li2015rare,li2018gapped}. Therefore, the magnetic ions should be almost homogeneously distributed in both diluted samples, otherwise significant Curie-Weiss temperatures should be expected. The difference between the Kramers Yb$^{3+}$ and non-Kramers Tm$^{3+}$ cases is clearly seen in $C_m$/$T$, where the signal of the diluted Yb$^{3+}$ sample diverges at low temperatures, whereas the diluted Tm$^{3+}$ sample reveals a finite zero-temperature value of $C_m$/$T$. This finite value indicates a distribution of the energy splitting between the two lowest-lying CEF singlets, $\mid$$E_1$$\rangle$ and $\mid$$E_2$$\rangle$~\cite{li2018gapped}. This transforms two singlets into a quasidoublet and gives rise to the Ising anisotropy~\cite{nekvasil1990effective,li2018gapped}. A similar single-ion scenario was recently reported for the Ising spin chain compound PrTiNbO$_6$~\cite{li2018gapped}. Here, we use the same approach and model the distribution with a Lorentzian function centered at $\overline{\Delta}$ = $\langle$$E_2$-$E_1$$\rangle$ and having the full width at half maximum (FWHM) $\omega$. The non-zero $\omega$ arises from the site mixing of Mg$^{2+}$ and Ga$^{3+}$ that, with their different charges, generate random CEF on the rare-earth site. By fitting $C_m$/$T$ of the diluted samples [Fig.~\ref{fig1} (b)], we find $\overline{\Delta}$ = 5.9 K and $\omega$ = 5.3 K for the Tm$^{3+}$ compound to be compared with $\overline{\Delta}$ = 0 K and $\omega$ = 0.19 K for Yb$^{3+}$, where the GS doublet is protected by time-reversal symmetry. Whereas this protection does not occur in the case of Tm$^{3+}$, a robust GS quasidoublet can still form, because $\omega$ is comparable to $\overline{\Delta}$.

Above 90 K, both the CEF randomness and intersite couplings, with the energy scales of $\sim$ 10 K [see Fig.~\ref{fig1} (a) and (b), see also below], can be neglected, and the combined CEF fit can be carried out for both magnetic susceptibilities and heat capacity measured on the single crystal of TmMgGaO$_4$ [see Fig.~\ref{fig1} (c) and (d)] (Appendix~\ref{a3}). Fitting thermodynamic data for Tm$_{0.04}$Lu$_{0.96}$MgGaO$_4$ leads to less accurate results, owing to the large error bar for the specific heat at high temperatures. Moreover, above $\sim$ 90 K the normalized susceptibilities for TmMgGaO$_4$ and Tm$_{0.04}$Lu$_{0.96}$MgGaO$_4$ nearly match, suggesting that intersite couplings play no significant role in this temperature range.

The average inner gap between the two lowest-lying CEF singlets is fitted to be ($E_2$-$E_1$)$^{CEF}$ = 6.3 K (Appendix~\ref{a3}), very similar to that in Tm$_{0.04}$Lu$_{0.96}$MgGaO$_4$, $\overline{\Delta}$ = 5.9 K (see above). $\Delta_{CEF}$ $\sim$ 450 K is also obtained for TmMgGaO$_4$, which is close to that of YbMgGaO$_4$~\cite{li2015rare,PhysRevLett.118.107202}. At low temperatures ($T$ $\ll$ $\Delta_{CEF}$), the components of the pseudospin-1/2 magnetic moment tensor are calculated as $m_{ij}^{\alpha}$ = $\mu_Bg_J\langle E_i|J_{\alpha}|E_j\rangle$ ($i$, $j$ = 1, 2, and $\alpha$ = $x$, $y$, $z$), where $g_J$ = 7/6 is the Land\'{e} $g$ factor and $J_{\alpha}$ is the component of the total angular momentum operator. And we obtain the general form of the tensor as
\begin{equation}
\mathbf{m^x}=\mathbf{m^y}=\left(
       \begin{array}{cc}
         0 & 0 \\
         0 & 0 \\
       \end{array}
     \right), \mathbf{m^z}=\frac{\mu_B}{2}\left(
       \begin{array}{cc}
         0 & G \\
         G^* & 0 \\
       \end{array}
     \right),
\label{Eq2}
\end{equation}
under the subspace of $|E_1\rangle$ and $|E_2\rangle$, where $|G|$ = $g_{\parallel}^{CEF}$ $\sim$ $g_{\parallel}$ [see Fig.~\ref{fig1} (a)]. The eigenstates of Eq.~(\ref{Eq2}) are $|\sigma=\pm\rangle$ = $\frac{1}{\sqrt{2}}(\frac{G}{|G|}|E_1\rangle\pm|E_2\rangle)$ with the Ising eigen-moments, $m^x$ = $m^y$ = 0 and $m^z$ $\sim$ $\pm$$\mu_Bg_{\parallel}$/2. Therefore, in the dipole approximation both $|E_1\rangle$ and $|E_2\rangle$ are nonmagnetic, but their linear superposition $|\sigma=\pm\rangle$ become magnetic. Under the subspace of $|\sigma=\pm\rangle$, we get
\begin{multline}
|E_1\rangle=\frac{1}{\sqrt{2}}\frac{|G|}{G}(|\sigma=+\rangle+|\sigma=-\rangle), \\
|E_2\rangle=\frac{1}{\sqrt{2}}(|\sigma=+\rangle-|\sigma=-\rangle).
\label{Eq3}
\end{multline}
Therefore, by resetting ($E_2$+$E_1$)/2 = 0 K the low-$T$ single-ion CEF term should be taken into account in the effective spin-1/2 Hamiltonian (see below),
\begin{equation}
\mathcal{H}_{single-ion}=\frac{\Delta}{2}(|E_2\rangle\langle E_2|-|E_1\rangle\langle E_1|) = -\Delta\cdot S_i^x,
\label{Eq4}
\end{equation}
with the random inner gap $\Delta$ = $E_2$-$E_1$. Different local environments at the Tm$^{3+}$ sites, with different distributions of Mg$^{2+}$/Ga$^{3+}$, give rise to the different CEF parameters, thus leading to different values of $\Delta$ as well as $g_{\parallel}$~\cite{PhysRevLett.118.107202}. Eq.~(\ref{Eq4}) introduces transverse magnetic field term into the spin Hamiltonian~\cite{moessner2000two}.

\section{Effective spin-1/2 Hamiltonian}

\begin{figure}[t]
\begin{center}
\includegraphics[width=8.8cm,angle=0]{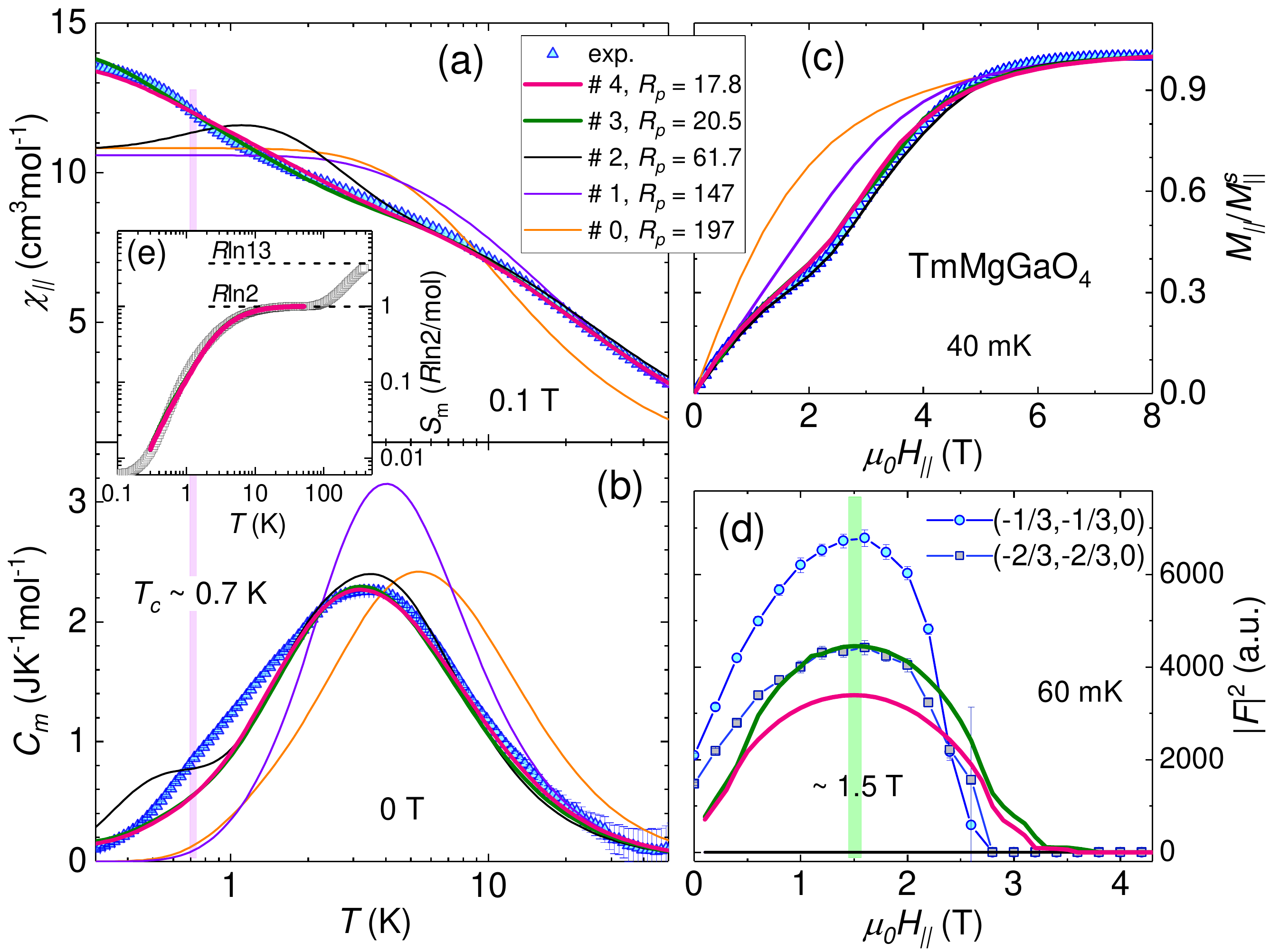}
\caption{(Color online)
Combined fits to the temperature dependence of (a) susceptibility and (b) magnetic heat capacity measured at $\sim$ 0 T, as well as (c) the field dependence of magnetization measured at 40 mK, on the single crystal of TmMgGaO$_4$, using the \#0, \#1, \#2, \#3, and \#4 models, respectively (see main text). (d) Magnetic field dependence of structure factors measured on the selected magnetic reflections at 60 mK on D23. The colored lines present the calculated values per Tm (multiplied by 1800 Tm, see Appendix~\ref{a5}) at 60 mK using the above \#1, \#2, \#3, and \#4 models. The \#1 and \#2 models give zero static structure factors on the magnetic reflections. (e) Magnetic entropy of TmMgGaO$_4$ measured at 0 T. The colored lines show the calculated values using the \#3 and \#4 models.}
\label{fig2}
\end{center}
\end{figure}

At $\sim$ 60 mK, $S_m$ of TmMgGaO$_4$ is measured to be nearly zero [see Fig.~\ref{fig2} (e)] suggesting that the system approaches its ground state. Coherent columnar magnetic reflections are indeed clearly observed by single-crystal neutron diffraction [see Fig.~\ref{fig1} (e)], with the fractional Miller indexes, $H$ = $\frac{n_1-n_2}{3}$ and $K$ = $\frac{n_1+2n_2}{3}$, where $n_1$ and $n_2$ are integers. The measured structural factors of these magnetic reflections are nearly independent on the third Miller index, $L$, at least from $L$ = -1.5 to 7 [Fig.~\ref{fig1} (f), and see Appendix~\ref{a5} for the linear plot]. We obtain a negligible interlayer correlation, $\xi_c$ $\sim$ 2$\pi$/FWHM$_L$ $<$ $c$/12, where FWHM$_L$ $>$ 12$\frac{2\pi}{c}$ is the broadening of the magnetic reflections along $L$~\cite{PhysRevB.70.214434,PhysRevB.88.024411}. Conversely, the crystal structure of TmMgGaO$_4$ is three-dimensional (3D), and the series of nuclear reflections are clearly observed with the integer Miller indexes [see Fig.~\ref{fig1} (f)], such as (2, 2, $L$) where $L$ = 0, 3, 6 ...; (2, 1, $L$) where $L$ = 1, 4 ...; (2, 0, $L$) where $L$ = -1, 2, 5 ...; and so on. Neutrons have a magnetic moment that is sensitive to the Tm$^{3+}$ dipole moments only. Therefore, our data directly evidence that an ideal 2D three-sublattice magnetic component of the dipole moments [see Fig.~\ref{fig1} (d)] forms in the 3D crystal structure of TmMgGaO$_4$ below 0.7 K and 2.6 T (see below), but the interlayer spin-spin correlations are negligible. A similar conclusion has been derived in Ref.~\cite{shen2018hidden} based on the time-of-flight inelastic neutron scattering (INS) data. To the best of our knowledge, experimental examples of ideal 2D magnetic structures in a real 3D material are highly rare to date, as the interlayer magnetic interactions are always present. The negligible interlayer couplings/correlations are likely caused by the extremely large interlayer distance of $c$/3 $\sim$ 8.4 \AA~in TmMgGaO$_4$~\cite{cevallos2017anisotropic}.

We can't assign unique values or detailed distributions (in the case of randomness) of $\langle S_1^z\rangle$, $\langle S_2^z\rangle$, $\langle S_3^z\rangle$, only based on the conventional magnetic structure refinement, because any three-sublattice structure illustrated in the inset of Fig.~\ref{fig1} (d) with arbitrary $\langle S_1^z\rangle$ $\neq$ $\langle S_2^z\rangle$ $\neq$ $\langle S_3^z\rangle$ gives the additional magnetic reflections sharing the equal structure factor, except for small differences [see Fig.~\ref{fig1} (f)] caused by the sample shape, crystal extinction, and similar effects~\cite{larson1994gsas}. Here, $\langle S_i^z\rangle$ = $\langle GS|S_i^z|GS\rangle$ and $|GS\rangle$ is the GS of the effective spin-1/2 system at $\sim$ 0 K, which is applicable to the following ED calculations. The neutron diffraction data do not contain enough information for the refinement of the 2D magnetic structure.

To explore the 2D correlated magnetism of TmMgGaO$_4$, as well as the detailed GS magnetic structures, one has to turn to the following disordered effective spin-1/2 Hamiltonian on the triangular lattice, with the longitudinal field of $H_{\parallel}$ applied along the $c$ axis,
\begin{multline}
\mathcal{H}=-\Delta\sum_iS_i^x+J_1^{zz}\sum_{\langle ij\rangle}S_i^zS_j^z+J_2^{zz}\sum_{\langle\langle ij'\rangle\rangle}S_i^zS_{j'}^z \\
-\mu_0H_{\parallel}\mu_Bg_{\parallel}\sum_iS_i^z.
\label{Eq5}
\end{multline}
Different from the 2D Ising Hamiltonian reported in Ref.~\cite{moessner2000two}, we should further consider the second-neighbor interaction in Eq.~(\ref{Eq5}), owing to the large observed $g$ factor, $g_{\parallel}$ = 13.18. In the limit of the magnetic dipole-dipole interaction, the average second-neighbor coupling is estimated to be non-negligible, $J_2^{zz}$ $\sim$ $\mu_0g_{\parallel}^2\mu_B^2$/(4$\pi r_{NNN}^3$) $\sim$ 0.53 K, where $r_{NNN}$ = $\sqrt{3}a$ = 5.9 \AA. The dipolar interaction is relatively long-ranged, and interactions beyond second neighbors can also be envisaged. However, the presence of these further-neighbor interaction terms significantly complicates the calculation and the modelling. In all of the existing references on TmMgGaO$_4$, other groups also restrict themselves to the second-neighbor interaction~\cite{bradley2019robust,shen2018hidden,li2019ghost}. Even if the couplings beyond second neighbors are non-negligible, we do not see strong reasons to include them into the effective spin Hamiltonian, in contrast to other perturbations, such as the randomness of the CEF gap (energy scale $\sim$ 8 K). And the present model of TmMgGaO$_4$ with only the first- and second-neighbor interactions should be just \emph{effective}. The real situation may be more complicated, but we seek to explain the bulk of experimental observations within the minimum model that captures the essential physics.

Between 30 and 60 K, $S_m$ is a constant of $R$ln2 [see Fig.~\ref{fig2} (e)], both the CEF excitations to higher levels and the intersite spin-spin correlations have marginal effect~\cite{li2018gapped}, and the mean-field approximation of the effective spin-1/2 system, $\chi_{\parallel}$ = $C_{\parallel}$/($T$-$\theta_w^{\parallel}$), is applicable. Here, $C_{\parallel}$ = $N_A\mu_0\mu_{eff}^2$/$k_B$ is the Curie constant, and the Curie-Weiss temperature of $\theta_w^{\parallel}$ = -3$\overline{(J_1^{zz}+J_2^{zz})}$/2 reflects the intersite magnetic couplings on the triangular lattice along the $c$ axis (in the spin space) ~\cite{li2015rare}. Through the Curie-Weiss fit to the susceptibility measured along the $c$ axis between 30 and 60 K, we obtain an effective moment of $\mu_{eff}$ = $g_{\parallel}\mu_{B}$/2 = 6.5(1)$\mu_{B}$ and $\theta_w^{\parallel}$ = -16.44(3) K. And we further get $\overline{J_1^{zz}+J_2^{zz}}$ $\sim$ -2$\theta_w^{\parallel}$/3 $\sim$ 10 K. In the above Curie-Weiss fit, we neglected the small van Vleck susceptibility (the CEF effect to the susceptibility), $\chi_{\parallel}^{vv}$ $<$ 0.2\%$\chi_{\parallel}$ at $T$ $\leq$ 60 K [see Fig.~\ref{fig1} (c)]. As $g_{\parallel}$ $\sim$ 2$J$$g_J$, $\mid$$m_J$=$\pm$$J$$\rangle$ dominate the GS CEF quasidoublet, and thus the non-Ising intersite coupling terms should be neglected in TmMgGaO$_4$~\cite{PhysRevLett.105.047201,PhysRevB.83.094411}.

\section{Fits to thermodynamic data}

\begin{figure*}[t]
\begin{center}
\includegraphics[width=17cm,angle=0]{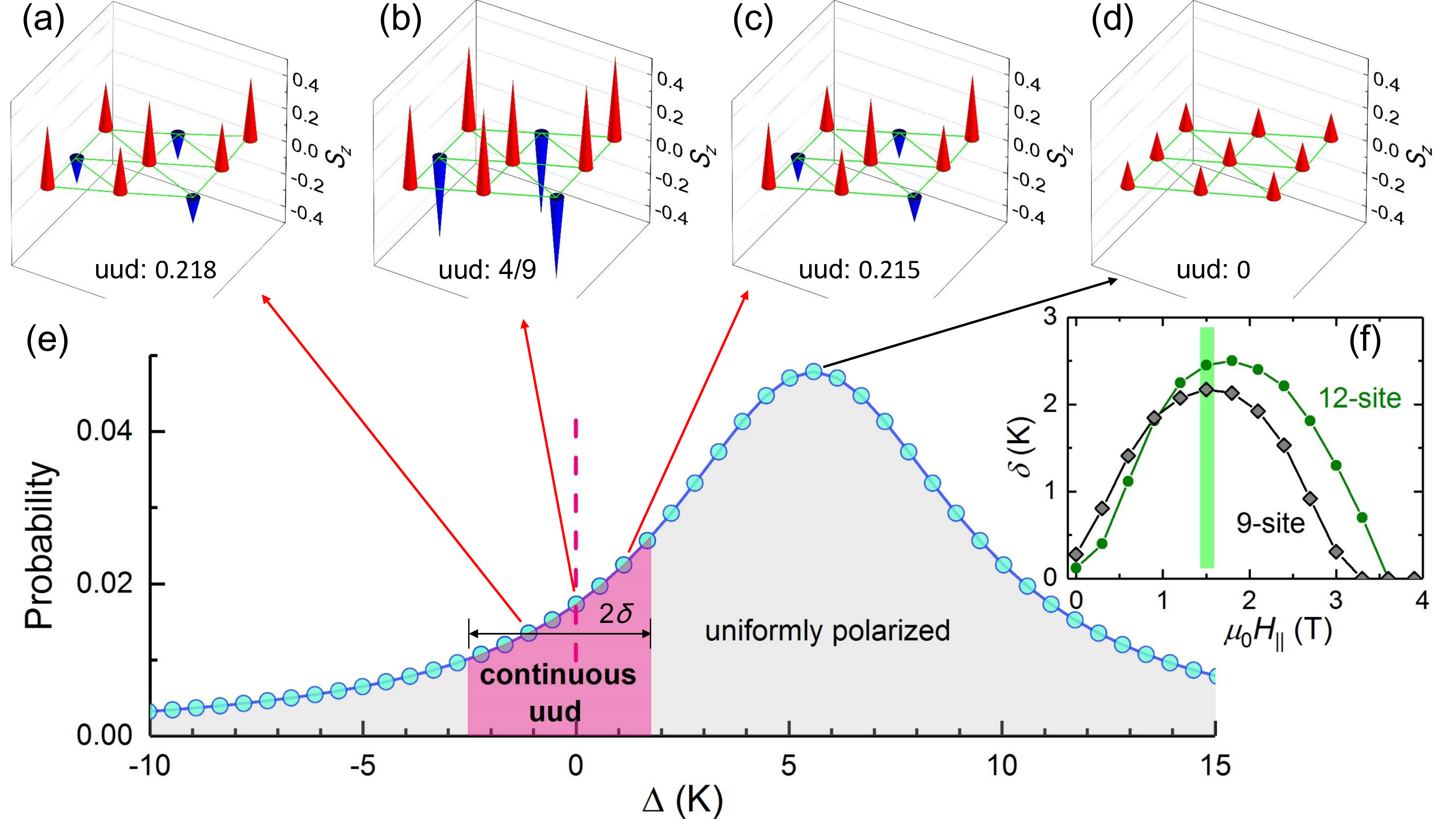}
\caption{(Color online)
2D continuous uud order in TmMgGaO$_4$ calculated by the least-$R_p$ optimized \#4 model under an external longitudinal field, $\mu_0H_{\parallel}$ = 1.5 T. The 9-site cluster with PBC is used in the ED calcualtion. The calculated (Ising) magnetic dipole structures at selected values of $\Delta$ = $E_2$-$E_1$ are shown in (a), (b), (c), and (d), respectively, with the numbers standing for the order parameter of the uud phase, as described in the text. (e) Lorentzian probability distribution of $\Delta$, $P$($\Delta$). A fraction ($\sim$ 11\%) of Tm$^{3+}$ ions within the range of 2$\delta$ (marked in red) give rise to the uud phase with the continuous distribution of the order parameter [see (a), (b), and (c) for example], while the remaining large fraction marked by gray results in the uniform polarization of the spin system [see (d) for example]. (f) Field dependence of $\delta$, $\delta$($H_{\parallel}$), calculated using the same Hamiltonian parameters of the fitted model \#4 (9-site ED) on the 9-site (black line) and 12-site (olive line) clusters with PBC.}
\label{fig3}
\end{center}
\end{figure*}

Since only the three-sublattice magnetic structure is observed by neutron diffraction below $T_c$ $\sim$ 0.7 K and below $\mu_0H_c$ $\sim$ 2.6 T, we carry out ED calculations using the 9-site and 12-site clusters with different periodic boundary conditions (PBC). No significant finite-size effects have been observed (Appendix~\ref{a4}). We use five different models: In \#1, we fix the parameters, $\Delta$ = 9.01 K, $J_1^{zz}$ = 6.61 K, $J_2^{zz}$ = 0.30 K, and $g_{\parallel}$ = 12.11, reported in Ref.~\cite{shen2018hidden} without any distribution, and get $R_p$ = 147. In \#2, we refine the above four parameters, and obtain $\Delta$ = 5.71(6) K, $J_1^{zz}$ = 10.9(1) K, $J_2^{zz}$ = 1.11(2) K, $g_{\parallel}$ = 13.6(1), and get the least-$R_p$ = 61.7~\footnote{Our result shows excellent agreement with the theoretical result reported in the recent preprint~\cite{li2019ghost}. Moreover, the differences between the ED calculations using the 9-site and 12-site clusters with different PBC are insignificant [see Fig.~\ref{fig3} (f) and Appendix~\ref{a4}].}. In \#3 and \#4, we further induce the Gaussian and Lorentzian distributions to $\Delta$, $g_{\parallel}$, $J_1^{zz}$, $J_2^{zz}$, respectively, due to the Mg/Ga site-mixing disorder. Each local chemical environment of Tm$^{3+}$ has a definite $\Delta$, and thus a definite $g_{\parallel}$~\cite{PhysRevLett.118.107202}. Moreover, the intersite couplings should also be distributed around their average values due to the CEF randomness, via both $f$-$p$ virtual electron hopping processes~\cite{PhysRevLett.105.047201,PhysRevB.83.094411,onoda2011effective} and magnetic dipole-dipole interactions ($\propto$ $g_{\parallel}^2$). Indeed, \#3 and \#4 fit the thermodynamic data much better, with much smaller least-$R_p$ = 20.5 and 17.8, respectively. \#3 gives $\overline{\Delta}$ = 5.66(6) K [FWHM = 12.8(2) K], $\overline{J_1^{zz}}$ = 8.57(8) K [FWHM = 1.13(2) K], $\overline{J_2^{zz}}$ = 2.36(3) K [FWHM = 2.19(4) K], and $\overline{g_{\parallel}}$ = 13.0(1) [FWHM = 0.93(1)]. And \#4 gives $\overline{\Delta}$ = 5.57(6) K [FWHM = 8.3(1) K], $\overline{J_1^{zz}}$ = 8.48(8) K [FWHM = 0.500(6) K], $\overline{J_2^{zz}}$ = 2.41(3) K [FWHM = 2.00(4) K], and $\overline{g_{\parallel}}$ = 13.0(1) [FWHM = 0.74(1)]. Finally, we also try to fit the thermodynamic data without any intersite couplings ($J_1^{zz}$ = $J_2^{zz}$ = 0, in the \#0 model), but the quality of the fit is very low with a large least $R_p$ = 197 (see Fig.~\ref{fig2}). Moreover, the four fitted parameters, $\overline{\Delta}$ = 10.5(1) K [FWHM = 13.1(2) K] and $\overline{g_{\parallel}}$ = 7.1(1) [FWHM = 11(1), 0 $\leq$ $g_{\parallel}$ $\leq$ 14], are inconsistent with the aforementioned values. Since only the single-ion terms are considered, the finite-size effects are completely excluded in this case. Therefore, we conclude that the low-$T$ physics of TmMgGaO$_4$ goes well beyond single-ion CEF effects, and the antiferromagnetic intersite couplings ($J_1^{zz}$ and $J_2^{zz}$) are critically important.

Besides the too large observed $R_p$ (see Fig.~\ref{fig2}), model \#1~\cite{shen2018hidden} can't well explain the measured magnetic properties of TmMgGaO$_4$ for the following reasons: First, the mean-field approximation, $\overline{J_1^{zz}}$+$\overline{J_2^{zz}}$ $\sim$ -2$\theta_w^{\parallel}$/3, must be fulfilled at high temperatures (30 $\leq$ $T$ $\leq$ 60 K). The least-$R_p$ fitted results of \#2, \#3, and \#4 obey the above relationship very well. In contrast, model \#1 gives $J_1^{zz}+J_2^{zz}$ = 6.9 K, much smaller than the reported value of -2$\theta_w$/3 = 12.7 K measured at 1 T in Ref.~\cite{shen2018hidden}, where this model was used. Second, the refined values of $\overline{\Delta}$ obtained from \#2, \#3, and \#4 models are much closer to 5.9 K measured on Tm$_{0.04}$Lu$_{0.96}$MgGaO$_4$ [Fig.~\ref{fig1} (b)], than to $\Delta$ = 9.01 K reported in Ref.~\cite{shen2018hidden}. Third, model \#1 essentially fails to reproduce the anomalies of the magnetization around $\mu_0H_c$ measured at low temperatures [see Fig.~\ref{fig2} (c)]. Fourth, models \#1 and \#2 without randomness completely miss the intensity increase of the magnetic reflections around $\mu_0H_{\parallel}$ $\sim$ 1.5 T measured at 60 mK [see Fig.~\ref{fig2} (d)]. In contrast, \#3 and \#4 models largely reproduce the above field dependence [see Fig.~\ref{fig2} (d)]. Therefore, the randomness caused by the Mg/Ga site mixing is an important ingredient to fully understand the novel low-$T$ correlated magnetism of TmMgGaO$_4$.

As the temperature increases, the zero-field integral intensities of the magnetic reflections gradually vanish at $T_c$ = 0.70(5) K, with showing a critical behavior (see Appendix~\ref{a5}). While, we did not observe any sharp peaks or anomalies in the temperature dependence of the magnetic susceptibility [Fig.~\ref{fig2} (a)] and heat capacity [Fig.~\ref{fig2} (b)], suggesting a short-range magnetic transition at $T_c$ in 0 T (see below). Therefore, we don't exclude from the fit any $T$-dependent data around $T_c$. And the deviation between the experimental data and least-$R_p$ \#3 (\#4) calculation is relatively large only in $C_m$ around $T_c$ [see Fig.~\ref{fig2} (b)]. Except this deviation, \#3 and \#4 models reproduce the entire field dependence of the magnetization measured at 40 mK [see Fig.~\ref{fig2} (c)], as well as the temperature dependence of the susceptibility [see Fig.~\ref{fig2} (a)]. Moreover, the entire magnetic entropy curve can be roughly reproduced by our models \#3 and \#4 below 50 K [see Fig.~\ref{fig2} (e)].

\section{Partial up-up-down order}

In TmMgGaO$_4$, the local chemical environments at the Tm$^{3+}$ sites lead to a distribution of the inner gap of the GS CEF quasidoublet, with a probability density function, $P(\Delta)$ [see Fig.~\ref{fig3} (e) for example]. At low temperatures, a large fraction of the Tm$^{3+}$ ions with the large inner gaps ($|\Delta|$ $>$ $\delta$) form the nonmagnetic component at $\mu_0H_{\parallel}$ $=$ 0 T, become uniformly polarized at $\mu_0H_{\parallel}$ $>$ 0 T [see Fig.~\ref{fig3} (d) for example], and thus can't contribute to the magnetic reflections with fractional $H$ and $K$.

On the other hand, the small fraction of the Tm$^{3+}$ ions with $|\Delta|$ $\leq$ $\delta$ contribute to the uud three-sublattice component [see Fig.~\ref{fig3} (a), (b), and (c) for example]. Here, we define the order parameter for this phase as $\frac{1}{N}\sum_i |\langle S_i^z\rangle-\frac{1}{N}\sum_i\langle S_i^z\rangle|$, where $N$ is the number of the triangular sites. The order parameter is strongly dependent on $\Delta$ and, therefore, continuously distributed. It takes the maximum of 4/9 at $\Delta$ = 0 K, and gradually vanishes at the boundaries, $|\Delta|$ $\sim$ $\delta$ (see Fig.~\ref{fig3}). Here, $\delta$($H_{\parallel}$) is strongly dependent on the applied longitudinal magnetic field [see Fig.~\ref{fig3} (f)]. It takes the maximum of $\sim$ 2 K at $\mu_0H_{\parallel}$ $\sim$ 1.5 T, which naturally explains the strongest magnetic neutron reflections observed at 60 mK and 1.5 T [Fig.~\ref{fig2} (d)]. The maximum distributed probability that gives rise to the uud order is only about 11\% observed at $\mu_0H_{\parallel}$ $\sim$ 1.5 T. Compared to the nuclear reflections, the magnetic ones with fractional $H$ and $K$, have much lower scattering intensities [see Fig.~\ref{fig1} (f)], which confirms the formation of only a small fraction of the uud order in TmMgGaO$_4$ at low temperatures. Quantitatively, the fully uud-ordered phase should give magnetic reflections with the intensity of $\sim$ 11000 (see Appendix~\ref{a4}) at $|\textbf{Q}|$ = 3.5276$\pi$/$a$, which is obviously larger than the average value of $\sim$ 1900 measured at 1.5 T (see Appendix~\ref{a5}). Therefore, the fraction of the uud compound can be estimated to be $\sim$ 1900/11000 $\sim$ 18\%, which is slightly larger than the above value of $\sim$ 11\% obtained from the thermodynamic data. In this case, the fraction should be in a range of $\sim$ 11$-$20\% at 1.5 T.

\begin{figure}[t]
\begin{center}
\includegraphics[width=8.8cm,angle=0]{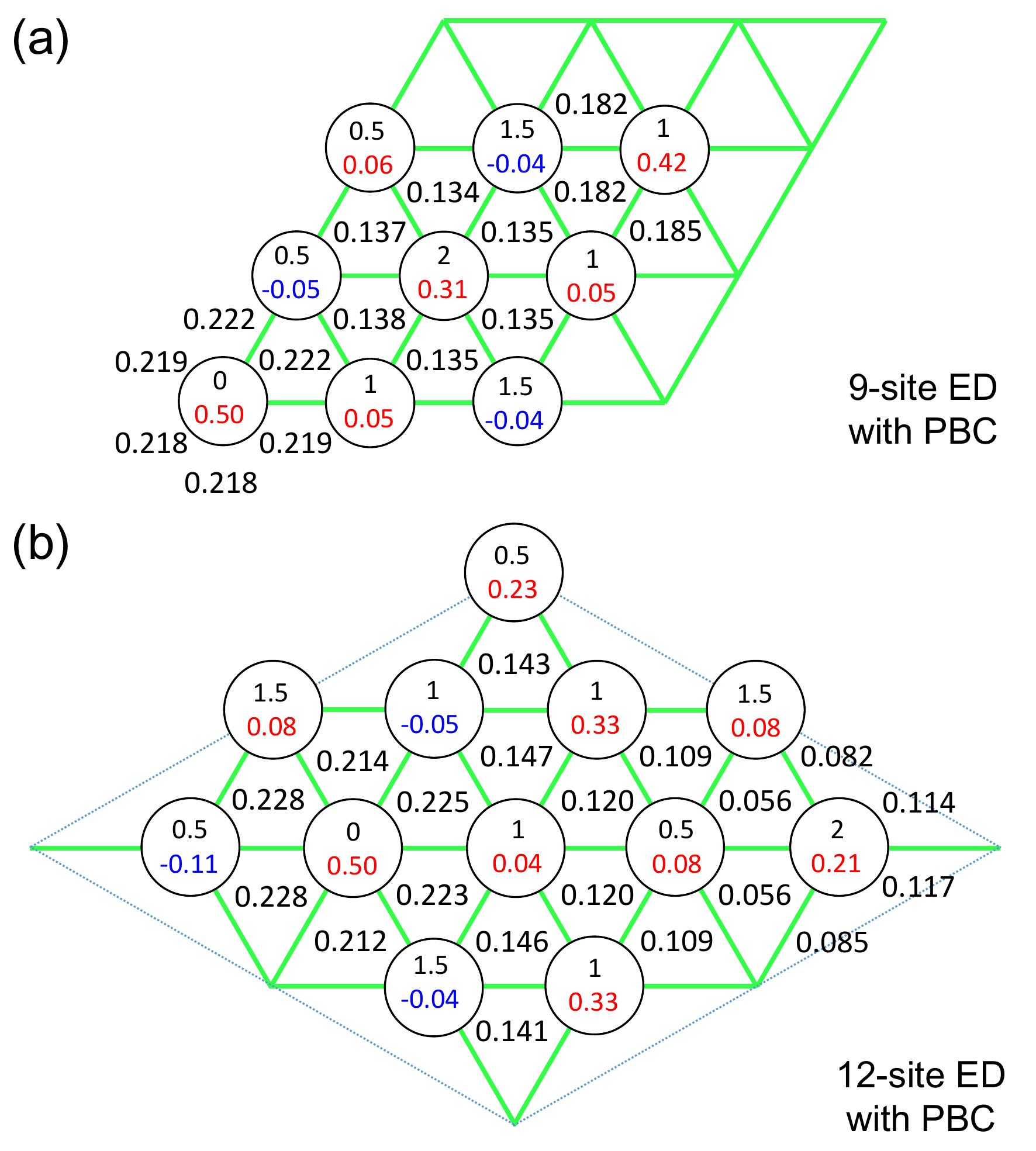}
\caption{(Color online)
Calculated ground state in the presence of the inhomogeneous randomness using the (a) 9-site and (b) 12-site clusters with different PBC, under $\mu_0H_{\parallel}$ = 1.5 T. The fitted parameters of the \#2 model ($\overline{\Delta}$ = 5.71 K, $J_1^{zz}$ = 10.9 K, $J_2^{zz}$ = 1.11 K, and $g_{\parallel}$ = 13.6) are used. The black number in the circle is the inner gap of each pseudospin (the multiple of $\overline{\Delta}$), and the red and blue numbers are the spin-up and spin-down static dipole moments, $\langle S_i^z\rangle$. The numbers in the triangles display the local order parameter calculated for the three spins at the corners of the respective triangle. The green lines depict the triangular lattice.}
\label{fig4}
\end{center}
\end{figure}

The above calculations are based on the hypothesis that local symmetries are preserved and same values of $\Delta$ occur within each cluster. However, the real situation in TmMgGaO$_4$ is much more complicated. Therefore, we also perform calculations in the presence of the spatially randomly distributed inner gap, $\Delta$. The average value of $\Delta$, $\overline{\Delta}$, is comparable to its FWHM, according to the magnetic heat capacity measured on the highly diluted sample of Tm$_{0.04}$Lu$_{0.96}$MgGaO$_4$ (see above). For simplicity, we assume that 1, 2, 3, 2, 1 pseudospins feature $\Delta$ = 0, 0.5$\overline{\Delta}$, $\overline{\Delta}$, 1.5$\overline{\Delta}$, 2$\overline{\Delta}$, respectively, in order to mimic the Lorentzian distribution. These pseudospins are randomly arranged on the 9-site cluster by Matlab [see Fig.~\ref{fig4} (a)]. Considering the possible size effect, we also performed the ED calculation on the 12-site cluster with PBC. Similarly, we assume that 1, 3, 4, 3, 1 pseudospins feature $\Delta$ = 0, 0.5$\overline{\Delta}$, $\overline{\Delta}$, 1.5$\overline{\Delta}$, 2$\overline{\Delta}$, respectively. The 12 pseudospins are randomly arranged in the cluster [see Fig.~\ref{fig4} (b)].

The calculations with the randomness inside the cluster largely confirm the main conclusions drawn from the previous calculations where local symmetries were kept. The local order parameter shows a pronounced variation (see Fig.~\ref{fig4}). Around the pseudospin with the smaller inner gap, the local order parameter is much larger than that around the pseudospin with the larger inner gap. Moreover, the uud and uniformly polarized components appear in different parts of the cluster depending on the local value of $\Delta$ [see Fig.~\ref{fig4} (b)]. These effects become even more obvious on larger clusters, but do not differ qualitatively from the results for the "homogeneous" clusters with same value of $\Delta$. Therefore, even the model without internal randomness within the cluster should capture the essential physics of TmMgGaO$_4$.

\section{Phase diagram and discussion}

\begin{figure}[t]
\begin{center}
\includegraphics[width=8.8cm,angle=0]{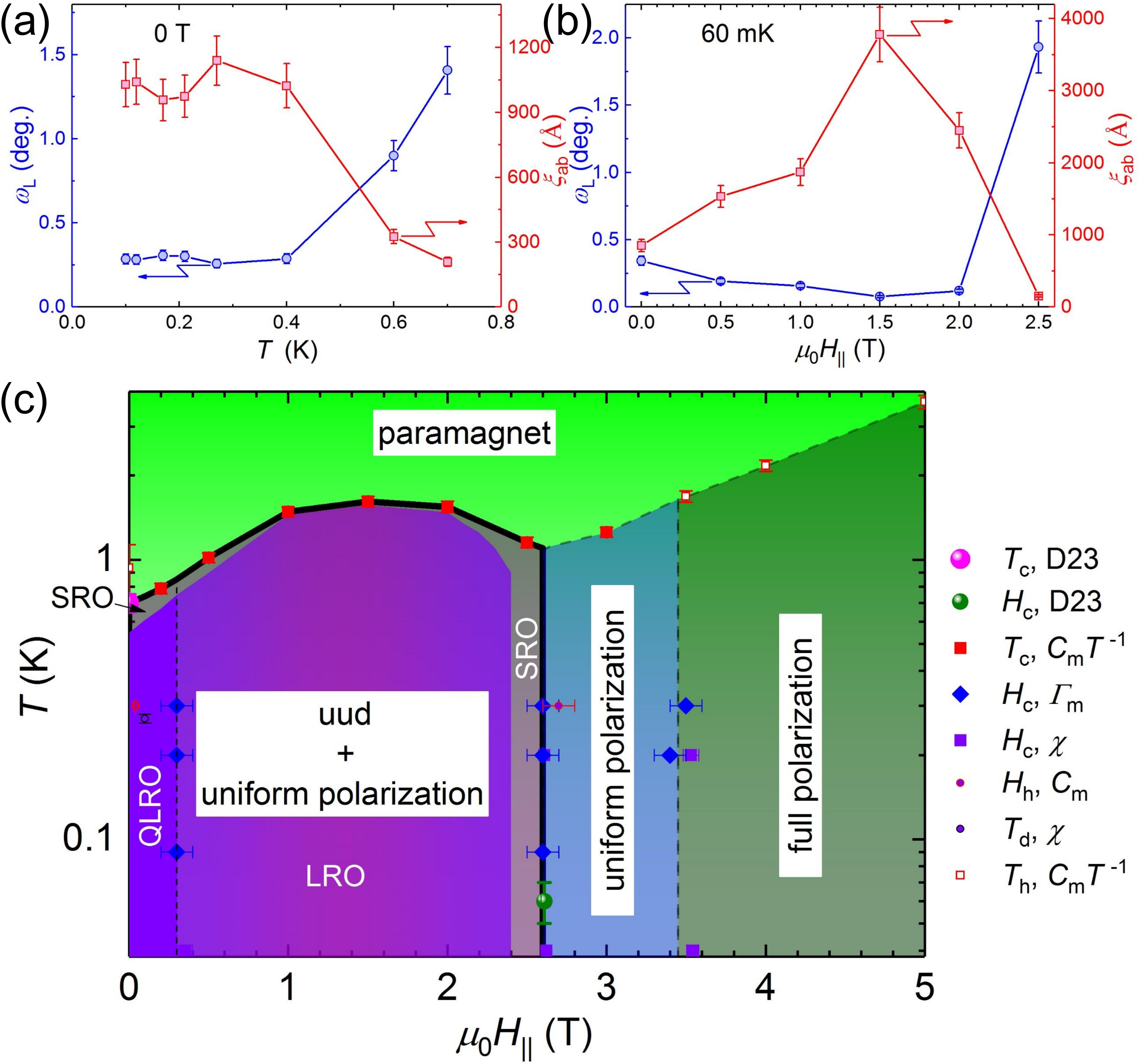}
\caption{(Color online)
(a) Temperature and (b) magnetic field dependence of the intrinsic reflection width (blue), $\omega_L$, as well as the in-plane correlation length (red), $\xi_{ab}$, extracted from the magnetic reflection, ($\frac{2}{3}$, -$\frac{1}{3}$, 0), measured on TmMgGaO$_4$ at 0 T and 60 mK, respectively. (c) Phase diagram of TmMgGaO$_4$ extracted from the neutron diffraction, heat capacity, magnetization, susceptibility, and magnetocaloric effect measurements. Long-range order (LRO, $\xi_{ab}$ $\geq$ 2000 {\AA}), quasi-long-range order (QLRO, $\xi_{ab}$ $\sim$ 1000 {\AA}), and short-range order (SRO, $\xi_{ab}$ $\leq$ 600 {\AA}) regions are marked roughly according to the magnetic neutron diffraction and heat capacity data.}
\label{fig5}
\end{center}
\end{figure}

Around the critical points, such as $T$ = $T_c$ at 0 T and $H_{\parallel}$ = $H_c$ at the low temperatures, the integral intensities of the magnetic reflections just completely disappear (see Appendix~\ref{a5}). The correlation length in the $ab$-plane ($\xi_{ab}$) can be estimated from the intrinsic broadening of the magnetic reflections along $H$ and $K$. Unlike the conventional long-range magnetic transition where the magnetic Bragg peaks keep coherent at all temperature below $T_c$~\cite{PhysRevB.70.214434}, TmMgGaO$_4$ shows diffuse magnetic Bragg peaks with short correlation lengths, $\xi_{ab}$ $\sim$ 200 {\AA} around $T_c$ and $H_c$ [see Fig.~\ref{fig5} (a) and (b)], well consistent with the absence of the sharp $\lambda$ peaks in the temperature dependence of the magnetic heat capacity at $T_c$. While, at the phase space well below the above critical points the magnetic Bragg peaks become sharp with (quasi-)long correlation lengths of $\geq$ 1000 {\AA}, which is more than two orders of magnitude larger than the lattice constant of the triangular lattice, $a$ = 3.4097 {\AA}. Moreover, the longitudinal magnetic field applied up to $\sim$ 1.5 T along the $c$ axis gradually shifts the critical point to a higher temperature, and the peak of the magnetic heat capacity becomes sharper and sharper [please see $C_m$/$T$ data in Fig.~\ref{figs2} (b) of Appendix~\ref{a2}]. These observations are in contradiction to the formation of the conventional short-range spin-glass GS~\cite{PhysRevLett.122.137201,PhysRevLett.110.137201,PhysRevLett.120.087201}.

To obtain the detailed low-$T$ phase diagram for TmMgGaO$_4$, we further measured temperature dependence of the magnetic heat capacity ($C_m$) at different magnetic fields, as well as the field dependence of $C_m$, the first derivative of magnetization (d$M_{\parallel}$/d$H_{\parallel}$), and magnetic Gr\"{u}neisen ratio ($\Gamma_m$) (see Fig.~\ref{figs2} in Appendix~\ref{a2}). The phase diagram is shown in Fig.~\ref{fig5} (c). Above $T_c$ = 0.70(5) K [see Fig.~\ref{figs7} (e) in Appendix~\ref{a5}], no magnetic neutron reflections with fractional $H$ and $K$ are observed, suggesting the paramagnetic phase with a large frustration factor, $|\theta_w^{\parallel}|$/$T_c$ $\sim$ 23. At low temperatures, the spin system of TmMgGaO$_4$ is fully polarized by high longitudinal fields, along with a very weak van Vleck susceptibility caused by excitations to higher-lying CEF levels. As the applied field decreases, both d$M_{\parallel}$/d$H_{\parallel}$ and $\Gamma_m$ show a broad hump at $\sim$ 3.5 T, indicating a crossover from the fully polarized phase to the uniformly partially polarized phase. No magnetic neutron reflections are observed above $\mu_0H_c$ = 2.61(2) T [see Fig.~\ref{figs7} (f) in Appendix~\ref{a5}], and both the optimized models \#3 and \#4 indeed produce $\delta$ = 0 K above $\sim$ 3 T, such that the uud order vanishes.

At $\mu_0H_c$, the emergence of the additional magnetic reflections clearly indicates the field-induced magnetic transition along with $\delta$ $>$ 0 K, confirmed by the relatively narrowed peak observed in the field dependence of d$M_{\parallel}$/d$H_{\parallel}$, $\Gamma_m$, and $C_m$. Narrowed peaks are observed at $\sim$ 1 K in the temperature dependence of $C_m$/$T$ at applied fields below $\mu_0H_c$, consistent with the phase transition toward the (quasi-)long-range uud order. At $\mu_0H_{\parallel}$ $\sim$ 1.5 T, the measured $C_m$/$T$ peak becomes sharpest and $\lambda$-shape at 1.61(7) K, and the magnetic neutron reflections take the maximum intensities with long-range correlations [$\xi_{ab}$ $\sim$ 4000 {\AA}, see Fig.~\ref{fig5} (b)] at 60 mK, which can be well interpreted by \#3 and \#4 models with the maximum $\delta$($H_{\parallel}$) [see Fig.~\ref{fig2} (d) and Fig.~\ref{fig3} (f)]. As $\mu_0H_{\parallel}$ further decreases, a broad peak is observed in the field dependence of d$M_{\parallel}$/d$H_{\parallel}$, $\Gamma_m$, and $C_m$, at $\sim$ 0.3 T, possibly suggesting some very delicate transition or crossover from the long-range ($\xi_{ab}$ $\sim$ 4000 {\AA}) to quasi-long-range ($\xi_{ab}$ $\sim$ 1000 {\AA}) orders along with the decrease of $\delta$($H_{\parallel}$). The above information of the correlation length is also roughly marked in the phase diagram [Fig.~\ref{fig5} (c)].

Despite the success of the models \#3 and \#4 in simultaneously reproducing the low-energy thermodynamic properties and magnetic order probed by neutron diffraction, we emphasize that a more sophisticated model would be required to describe all experimental data, including the spin-wave excitations reported in Ref.~\cite{shen2018hidden}. First, the asymmetry of the peak is clearly observed in the field dependence of the structure factor on the magnetic reflection measured at 60 mK [see Fig.~\ref{fig2} (d)]. There is still about 30\% of the maximum intensity (at 1.5 T) observed at 0 T, and the structure factor quickly disappears at $\mu_0H_c$ = 2.61(2) T. In contrast, both models \#3 and \#4 give the symmetric peak profile centered at $\sim$ 1.5 T [see Fig.~\ref{fig2} (d)]. Second, the calculated INS excitations using \#3 and \#4 indeed get broader, but may still deviate from the reported spin-wave result~\cite{shen2018hidden} (see Appendix~\ref{a4}). The asymmetric distribution functions of $\Delta$, $g_{\parallel}$, $J_1^{zz}$, and $J_2^{zz}$ by considering the detailed Mg/Ga arrangements, as well as the inherent correlations among these Hamiltonian parameters, would be required to reach a consistent interpretation for all observations.

A similar effective spin-1/2 Ising Hamilitonian with a continuous distribution of the microscopic parameters can be applied to other non-Kramers rare-earth magnets with correlated GS quasidoublets, such as the Pr$^{3+}$ effective spin-1/2 chain compound, PrTiNbO$_6$, with the similar site-mixing disorder between nonmagnetic Ti$^{4+}$ and Nb$^{5+}$ ions~\cite{li2018gapped}. Therefore, the correlated magnetism of the GS quasidoublets should be general as well in condensed matter physics, as the structural disorder is usually inevitable in a real material. Our present work paves the road to understanding this kind of novel many-body physics.

\section{Conclusions}

We performed an extensive single-crystal study on the 2D frustrated magnetism of TmMgGaO$_4$, with the perfect triangular lattice of non-Kramers rare-earth Tm$^{3+}$ ions. The distribution of two nearly degenerate GS CEF singlets (quasidoublet) caused by the Mg/Ga disorder is clearly evidenced by the magnetic heat capacity of highly diluted Tm$_{0.04}$Lu$_{0.94}$MgGaO$_4$ with the negligible intersite couplings, as well as the combined CEF fits to the high-$T$ thermodynamic data of TmMgGaO$_4$. At low temperatures, the effective spin-1/2 Hamiltonian of the correlated quasidoublets is experimentally determined. It gives rise to the small fraction of the 2D uud phase of the Ising magnetic dipoles with small inner gaps ($|\Delta|$ $\leq$ $\delta$), as well as the main nonmagnetic phase at 0 T with large inner gaps ($|\Delta|$ $>$ $\delta$), which become uniformly polarized at a finite longitudinal applied field of $H_{\parallel}$. Our correlated quasidoublet model naturally explains the strongest magnetic reflections observed at $\mu_0H_{\parallel}$ $\sim$ 1.5 T, as well as the vanishing intensity with increasing or decreasing $H_{\parallel}$. The similar effective spin-1/2 model with a distribution of the microscopic parameters should be applied to other non-Kramers rare-earth magnets with the disorder-induced GS CEF quasidoublets.

\acknowledgements

We thank Eric Ressouche and Pascal Fouilloux for their technical help at the ILL, as well as Sebastian Esser for his technical help in G\"{o}ttingen, and Yuanpai Zhou for his technical help with the calculations. Y. L. thanks Wei Li and Ke Liu for the helpful discussion, and Prof. Liang Li for the invitation for the academic visit at Huazhong University of Science and Technology. Y. L. was supported by the start-up fund of Huazhong University of Science and Technology in China. The work in Augsburg was supported by the German Science Foundation via the project 107745057 (TRR80) and by the German Federal Ministry for Education and Research through the Sofja Kovalevskaya Award of the Alexander von Humboldt Foundation.

\section*{Appendix}

\appendix

\section{Sample synthesis and characterization above 1.8 K}
\label{a1}

\begin{figure}[t]
\begin{center}
\includegraphics[width=8.6cm,angle=0]{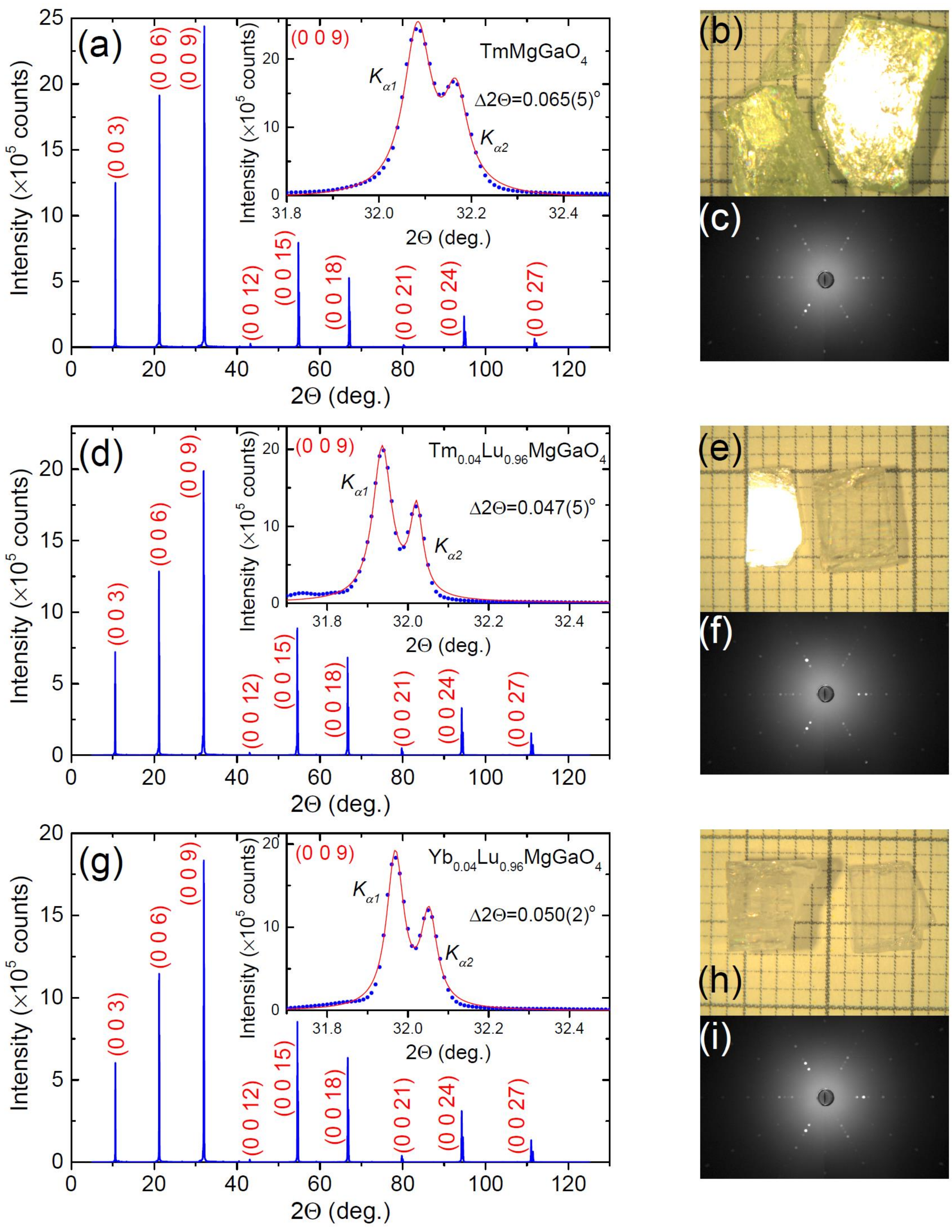}
\caption{(Color online)
(a) X-ray diffraction for the TmMgGaO$_4$ single crystal on the $ab$-plane. The inset presents a zoom-in plot of the strongest Bragg peak, (0 0 9), where the angle (2$\Theta$) difference between the nearest-neighbor data points is 0.01$^o$. (b) Single crystals of TmMgGaO$_4$ cut along the $ab$-plane. (c) Laue x-ray diffraction pattern of TmMgGaO$_4$ on the $ab$-plane. (d) X-ray diffraction for the Tm$_{0.04}$Lu$_{0.96}$MgGaO$_4$ single crystal on the $ab$-plane. The inset presents a zoom-in plot of the strongest Bragg peak, (0 0 9). (e) Single crystals of Tm$_{0.04}$Lu$_{0.96}$MgGaO$_4$ cut along the $ab$-plane. (f) Laue x-ray diffraction pattern of Tm$_{0.04}$Lu$_{0.96}$MgGaO$_4$ on the $ab$-plane. (g) X-ray diffraction for the Yb$_{0.04}$Lu$_{0.96}$MgGaO$_4$ single crystal on the $ab$-plane. The inset presents a zoom-in plot of the strongest Bragg peak, (0 0 9). (h) Single crystals of Yb$_{0.04}$Lu$_{0.96}$MgGaO$_4$ cut along the $ab$-plane. (i) Laue x-ray diffraction pattern of Yb$_{0.04}$Lu$_{0.96}$MgGaO$_4$ on the $ab$-plane.}
\label{figs1}
\end{center}
\end{figure}

\begin{figure}[t]
\centering
\includegraphics[width=8.6cm,angle=0]{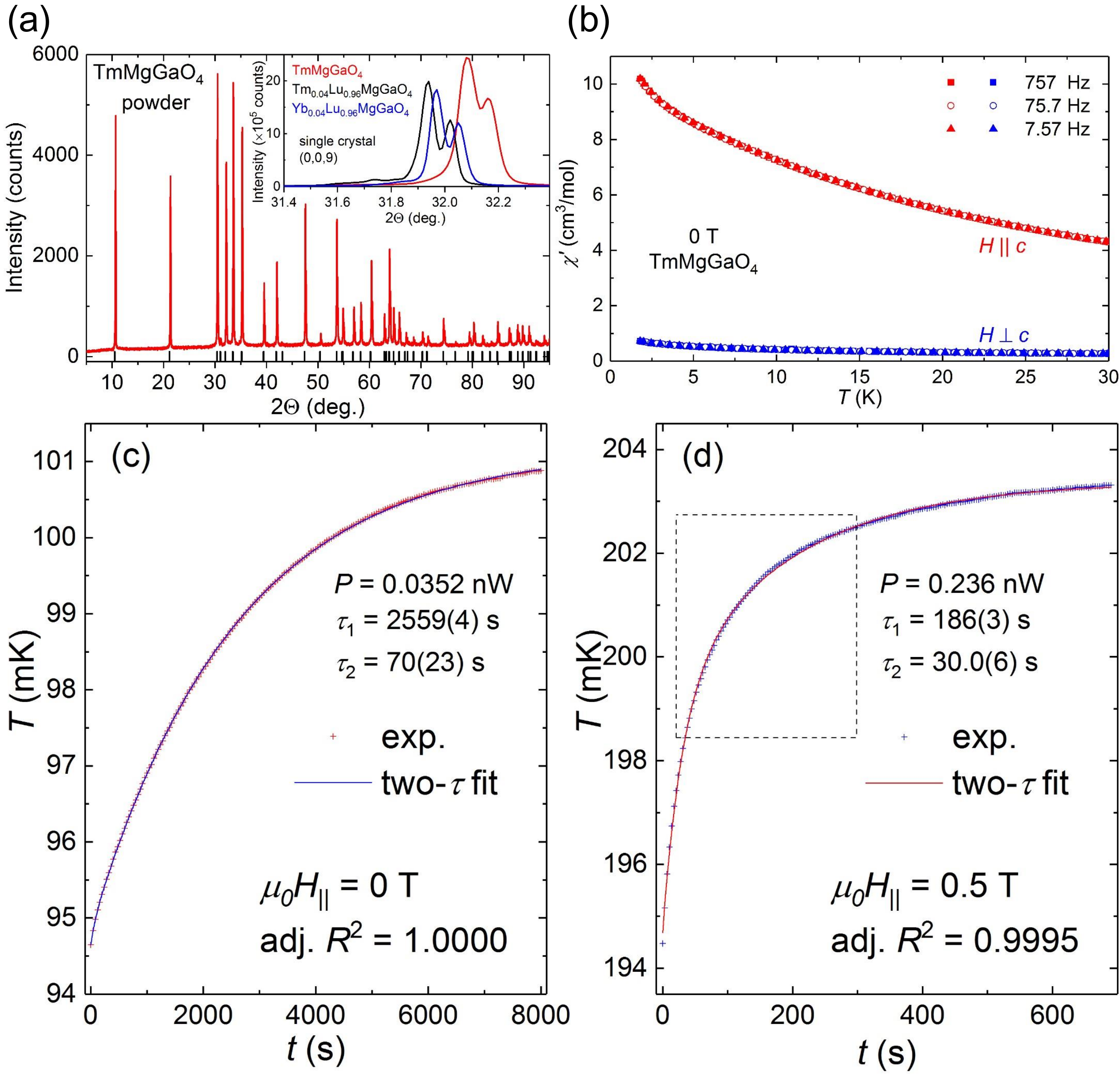}
\caption{(Color online)
(a) Powder x-ray diffraction measured on the polycrystalline sample of TmMgGaO$_4$ at 300 K. The black bars show the reflections calculated with the crystal structure data reported by Cevallos \emph{et al.}~\cite{cevallos2017anisotropic}. The inset shows the strongest Bragg peaks, (0, 0, 9), measured on the single crystals of TmMgGaO$_4$, Tm$_{0.04}$Lu$_{0.96}$MgGaO$_4$, and Yb$_{0.04}$Lu$_{0.96}$MgGaO$_4$, respectively. (b) Temperature dependence of the ac susceptibility (the real part) measured on the single crystal of TmMgGaO4 down to 1.8 K. Thermal relaxation data of TmMgGaO$_4$ single crystal measured (c) at 0 T at $\sim$ 0.1 K, (d) at 0.5 T at $\sim$ 0.2 K, with the lines representing the least-square fits using the two-$\tau$ model~\cite{li2018gapped}. For the definition of adj. $R^2$, please see https://www.originlab.com/doc/Origin-Help/Interpret-Regression-Result.}
\label{figs_r1}
\end{figure}

Large and transparent single crystals ($\sim$ 1 cm) of TmMgGaO$_4$, Tm$_{0.04}$Lu$_{0.96}$MgGaO$_4$, and Yb$_{0.04}$Lu$_{0.96}$MgGaO$_4$ (See Fig.~\ref{figs1}) were grown in a high-temperature optical floating zone furnace (FZ-T-10000-H-VI-VPM-PC, Crystal Systems Corp.), using 53.0\%, 60.7\%, and 60.9\% of the full power of the four lamps (the full power is 1.5 kW for each lamp), respectively~\cite{li2015rare,cevallos2017anisotropic,li2018gapped}. The single crystals were oriented by the Laue x-ray diffraction, and were cut consequently by a line cutter along the crystallographic $ab$ plane. The cut planes were cross-checked by both Laue (see Fig.~\ref{figs1}) and conventional x-ray diffraction (see Fig.~\ref{figs1}). The high-quality of the crystal was confirmed by the narrow reflection peaks, 2$\Delta\Theta$ = 0.047$-$0.065$^{o}$ (full width at half maximum, FWHM). The un-indexed broad hump at $\sim$ 31.74$^o$ possibly comes from the tape used for measuring the Tm$_{0.04}$Lu$_{0.96}$MgGaO$_4$ crystal. Because we fixed the surface of the crystal on a piece of tape, broad humps with the width of $\sim$ 0.3$^o$ may be sometimes detected. No such features are observed on the single crystals of TmMgGaO$_4$ and Yb$_{0.04}$Lu$_{0.96}$MgGaO$_4$ [see Fig.~\ref{figs_r1} (a)].

No significant impurity phase of the TmMgGaO$_4$ sample was observed by the single-crystal x-ray and neutron diffraction, consistent with the previously reported work~\cite{cevallos2017anisotropic}. These single-crystal samples are well transparent, and thus we have full confidence in the absence of any impurity phases, also from a visual inspection of the crystals with a microscope. We also show the x-ray diffraction data measured on the TmMgGaO$_4$ powder in Fig.~\ref{figs_r1} (a), confirming no obvious impurity phase.

The dc magnetization (1.8 $\leq$ $T$ $\leq$ 400 K and 0 $\leq$ $\mu_0H$ $\leq$ 7 T) was measured by a magnetic property measurement system (MPMS, Quantum Design) using single crystals of $\sim$ 100 mg. The dc magnetization up to 14 T was measured by a vibrating sample magnetometer (VSM) in a physical property measurement system (PPMS, Quantum Design). The heat capacity (1.8 $\leq$ $T$ $\leq$ 400 K and 0 $\leq$ $\mu_0H$ $\leq$ 12 T) was measured using single crystals of $\sim$ 10 mg in a PPMS. N-grease was used to facilitate thermal contact between the sample and the puck below 210 K, while H-grease was used above 200 K. The sample coupling was better than 99\%. The contributions of the grease and puck under different external fields were measured independently and subtracted from the data. It is very difficult to precisely measure the magnetic heat capacity of Tm$_{0.04}$Lu$_{0.96}$MgGaO$_4$ and Yb$_{0.04}$Lu$_{0.96}$MgGaO$_4$ above 10 K, due to the high dilution of the magnetic ions and the inevitable thermal disturbance.

The ac susceptibility (1.8 $\leq$ $T$ $\leq$ 30 K and 0 T) was measured by the MPMS using a single crystal of TmMgGaO$_4$ with a mass of $\sim$ 100 mg [see Fig.~\ref{figs_r1} (b)]. And no obvious frequency dependence was observed from 7.57 to 757 Hz down to 1.8 K.

\section{Millikelvin measurements below 2 K.}
\label{a2}

\begin{figure}[t]
\centering
\includegraphics[width=8.6cm,angle=0]{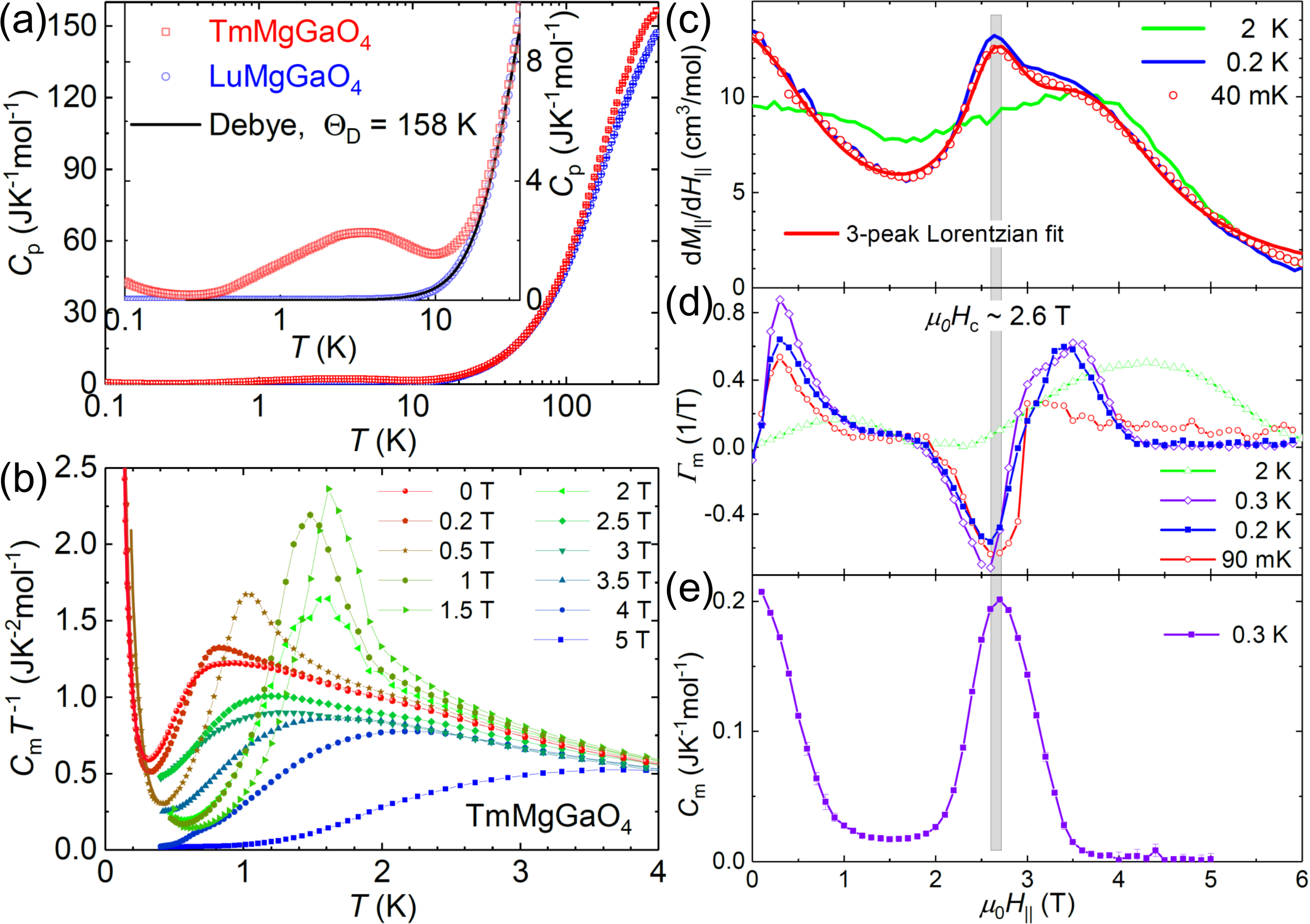}
\caption{(Color online)
(a) Heat capacity of the TmMgGaO$_4$ and LuMgGaO$_4$ single crystals measured at 0 T. The inset presents a zoom-in plot of the low-$T$ data with the black line showing the Debye heat-capacity fit ($\Theta_D$ = 158 K). (b) Magnetic heat capacity of TmMgGaO$_4$ measured at selected fields. The phonon or lattice contribution was subtracted by the heat capacity of the non-magnetic LuMgGaO$_4$. The observed upturns below $\sim$ 0.3 K are fitted by considering the nuclear spin contributions. (c) Field dependence of the susceptibility (d$M_{\parallel}$/d$H_{\parallel}$) with the red line showing the three-peak Lorentzian fit. (d) Field dependence of the magnetic Gr\"{u}neisen ratio measured at 0.09, 0.2, 0.3, and 2 K. (e) Field dependence of the heat capacity measured at 0.3 K.}
\label{figs2}
\end{figure}

The total heat capacity ($C_p$) of the TmMgGaO$_4$, LuMgGaO$_4$, Tm$_{0.04}$Lu$_{0.96}$MgGaO$_4$, and Yb$_{0.04}$Lu$_{0.96}$MgGaO$_4$ single crystals was measured by a home-built setup in a $^3$He-$^4$He dilution refrigerator between 0.1 and 2.0 K at magnetic fields up to 5 T applied along the $c$ axis. In contrast to the commercial PPMS, both the thermal link and thermometer are directly attached to the upper surface of the single-crystal sample with the well-polished bottom surface, which is attached to the upper surface of the platform using grease. The heater is mounted on the bottom surface of the platform~\cite{li2018gapped}. The two-$\tau$ model~\cite{li2018gapped} is applicable in most cases. No signatures of the poor thermal contact between the sample and holder were observed.

\begin{figure}[t]
\centering
\includegraphics[width=8.6cm,angle=0]{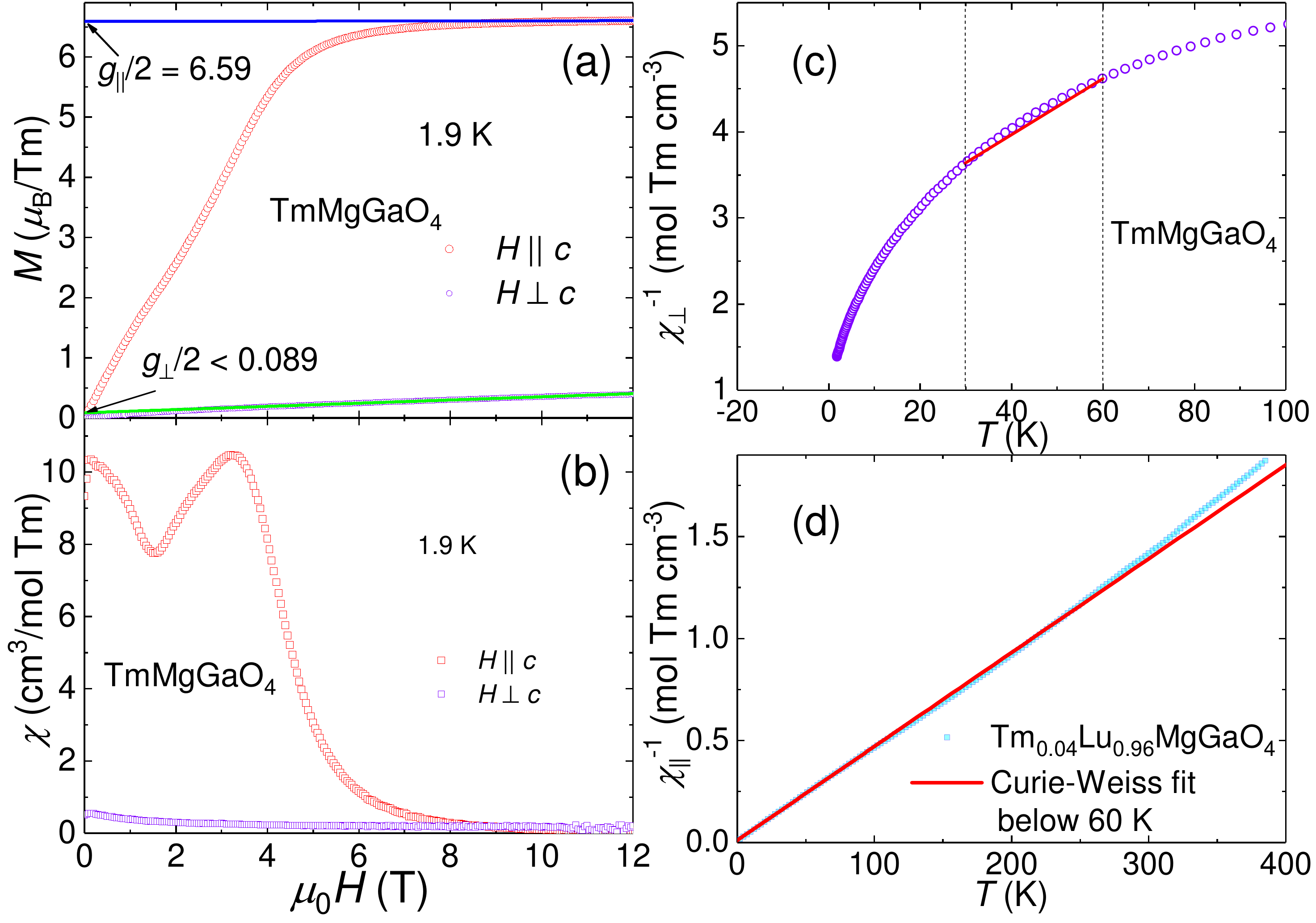}
\caption{(Color online)
Field dependence of (a) magnetization ($M$) and (b) susceptibility (d$M$/d$H$) of TmMgGaO$_4$ measured in fields both parallel and perpendicular to the $c$-axis. The data were measured in the same VSM-PPMS using the same single crystal of TmMgGaO$_4$ (96.90 mg). The colored lines in (a) represent the linear fits above 8 T. We stopped the $M_{\perp}$ measurement at 12 T, as the force acting on the crystal may become too large, thus breaking the crystal itself or the sample holder~\cite{li2018gapped}. (c) Temperature dependence of the susceptibility measured in the field of 0.1 T applied perpendicular to the $c$ axis. The straight red line is a guide to the eye demonstrating that no Curie-Weiss behavior is observed between 30 and 60 K [the temperature range where the magnetic entropy in Fig.~\ref{fig2} (e) of the main text shows a plateau that indicates a paramagnetic regime not affected by spin-spin correlations and CEF excitations to higher levels]. (d) Temperature dependence of the dc susceptibility measured on the single crystal of Tm$_{0.04}$Lu$_{0.96}$MgGaO$_4$ along the $c$ axis. The red line shows the Curie-Weiss fit to the data between 30 and 60 K.}
\label{fig_r3}
\end{figure}

In Fig.~\ref{figs_r1} (c), we show a typical zero-field relaxation curve and its fit with the two-$\tau$ model. In applied magnetic fields, the data may deviate from the two-$\tau$ model at short times even at higher temperatures [as in Fig.~\ref{figs_r1} (d)], similar to our previous report on the spin-chain compound PrTiNbO$_6$~\cite{li2018gapped}. This slight deviation may be caused by the thermal decoupling between the phonon (lattice) and electronic/nuclear subsystems~\cite{smith2005origin}. Similar to this previous work~\cite{li2018gapped}, we chose to exclude the heat capacity data with the adj. $R^2$ smaller than 0.9995.

The magnetic heat capacity ($C_m$) of TmMgGaO$_4$ was obtained by subtracting $C_p$ of LuMgGaO$_4$ from $C_p$ of TmMgGaO$_4$ [see Fig.~\ref{figs2} (a)]. We fitted the 0, 0.2, and 0.5 T heat capacities using the function, $C_n$($^{169}\Delta$/$T$)+$A$exp(-$\Delta$/$T$), from the lowest temperature up to the temperature of the minimum in $C_m$/$T$ [see Fig.~\ref{figs2} (b)]. Here $C_n$($^{169}\Delta$/$T$) is the nuclear heat capacity expressed by a two-level model, $^{169}\Delta$ and $\Delta$ are the nuclear and electronic spin gaps, respectively, and $A$ is a pre-factor~\cite{li2018gapped}.

The dc magnetization ($M_{\parallel}$) of TmMgGaO$_4$ between 0.024 and 2.0 K at magnetic fields up to 8 T applied along the $c$ axis, was measured by a high-resolution capacitive Faraday force magnetometer in a $^3$He-$^4$He dilution refrigerator~\cite{sakakibara1994faraday} [see Fig.~\ref{figs2} (c) for d$M_{\parallel}$/d$H_{\parallel}$]. The magnetic Gr\"{u}neisen ratio or magnetocaloric effect, $\Gamma_m$ = ($dT$/$dH$)/($\mu_0T$) = -($dM_{\parallel}$/$dT$)/$C_p$, was measured by the alternating field technique ($\nu$ = 0.02 and 0.04 Hz) in a $^3$He-$^4$He dilution refrigerator~\cite{tokiwa2011high,tokiwa2014quantum} [see Fig.~\ref{figs2} (d)]. Field dependence of the heat capacity was also measured at 0.3 K [see Fig.~\ref{figs2} (e)], where both the nuclear and lattice contributions are negligible.  The peak positions correspond to the magnetic field induced transitions or crossovers (see Fig.~\ref{figs2}).

\section{Combined CEF fit for TmMgGaO$_4$.}
\label{a3}

The Ising nature of the Tm$^{3+}$ magnetic moments is evidenced by the strongly anisotropic magnetization (Fig.~\ref{fig_r3}). With $H\parallel c$, the magnetization becomes linearly field-dependent above 8 T, with the intercept of $g_{\parallel}$/2, where $g_{\parallel}$ $\sim$ 13.18 [see Fig.~\ref{fig_r3} (a)]. In contrast, the magnetization measured in the field perpendicular to $c$, remains very low and corresponds to the nearly field-independent susceptibility $dM$/$dH$ [see Fig.~\ref{fig_r3} (b)]. This susceptibility is mostly of the van Vleck origin, whereas for the non-van Vleck part we can put the upper limit of $g_{\perp}$ $<$ 0.18 and estimate $g_{\perp}$/$g_{\parallel}$ $<$ 1.4\%, as the slightly tilting of the sample by the applied field of $H_{\perp}$ is enough to account for the observed weak non-van Vleck part.

For $H\perp c$, we observed a non-Curie-Weiss temperature-dependent behavior in $\chi_{\perp}$ between 30 and 60 K [Fig.~\ref{fig_r3} (c)], as expected in the presence of the dominant van Vleck term.

The Curie-Weiss behavior of Tm$_{0.04}$Lu$_{0.96}$MgGaO$_4$ with a very small $\theta_w^{\parallel}$ (fitted below 60 K) extends up to $\sim$ 100 K [Fig.~\ref{fig_r3} (d)], which also confirms the low-$T$ effective spin-1/2 physics and the neglectable CEF effect on $\theta_w^{\parallel}$ between 30 and 60 K. The deviation from the Curie-Weiss law above $\sim$ 100 K should be caused by excitations to higher CEF levels, and that is the CEF effect to the susceptibility (see below in Fig.~\ref{CEF}).

\begin{table}[t]
\caption{CEF parameters, $B_{n}^{m}$, obtained from the combined fit. The units are in meV.}\label{table1}
\begin{center}
\begin{tabular}{ l | l | l | l | l | l }
    \hline
    \hline
 $B_{2}^{0}$ & $B_{4}^{0}$ & $B_{4}^{3}$ & $B_{6}^{0}$ & $B_{6}^{3}$ & $B_{6}^{6}$ \\ \hline
 -0.58 & -0.00068 & -0.036 & -0.0000047 & -0.00123 & -0.0000178 \\
    \hline
    \hline
\end{tabular}
\end{center}
\end{table}

\begin{table}[t]
\caption{Fitted CEF energy levels and the corresponding CEF states under 0 T.}\label{table2}
\begin{center}
\begin{tabular}{ l }
    \hline
    \hline
 $E_1$ = 0 K \\
 $|E_1\rangle$ = 0.63($|6\rangle$+$|$-6$\rangle$)+0.32($|3\rangle$-$|$-3$\rangle$)-0.10$|0\rangle$ \\ \hline
 $E_2$ = 6.3 K \\
 $|E_2\rangle$ = 0.63($|6\rangle$-$|$-6$\rangle$)+0.32($|3\rangle$+$|$-3$\rangle$) \\ \hline
 $E_3$ or $E_4$ = 446 K \\
 $|E_3\rangle$ = 0.91$|5\rangle$+0.42$|2\rangle$-0.05$|$-1$\rangle$ \\
 $|E_4\rangle$ = 0.91$|$-5$\rangle$-0.42$|$-2$\rangle$-0.05$|$1$\rangle$ \\ \hline
 $E_5$ or $E_6$ = 702 K \\
 $|E_5\rangle$ = 0.98$|4\rangle$-0.19$|$1$\rangle$ \\
 $|E_6\rangle$ = 0.98$|$-4$\rangle$+0.19$|$-1$\rangle$ \\ \hline
 $E_7$ = 810 K \\
 $|E_7\rangle$ = 0.26($|6\rangle$+$|$-6$\rangle$)-0.39($|$3$\rangle$-$|$-3$\rangle$)+0.74$|0\rangle$ \\ \hline
 $E_8$ or $E_9$ = 905 K \\
 $|E_8\rangle$ = 0.30$|$5$\rangle$-0.17$|$-4$\rangle$-0.54$|$2$\rangle$+0.75$|$-1$\rangle$ \\
 $|E_9\rangle$ = 0.30$|$-5$\rangle$+0.17$|4\rangle$+0.54$|$-2$\rangle$+0.75$|1\rangle$ \\ \hline
 $E_{10}$ = 1014 K \\
 $|E_{10}\rangle$ = 0.32($|6\rangle$-$|$-6$\rangle$)-0.63($|$3$\rangle$+$|$-3$\rangle$) \\
 \hline
 $E_{11}$ or $E_{12}$ = 1047 K \\
 $|E_{11}\rangle$ = 0.29$|$5$\rangle$+0.10$|$-4$\rangle$-0.72$|$2$\rangle$-0.61$|$-1$\rangle$ \\
 $|E_{12}\rangle$ = 0.29$|$-5$\rangle$-0.10$|4\rangle$+0.72$|$-2$\rangle$-0.61$|1\rangle$
 \\ \hline
 $E_{13}$ = 1192 K \\
 $|E_{13}\rangle$ = 0.20($|6\rangle$+$|$-6$\rangle$)-0.49($|$3$\rangle$-$|$-3$\rangle$)-0.66$|0\rangle$ \\
    \hline
    \hline
\end{tabular}
\end{center}
\end{table}

The strict Ising anisotropy is possibly related to the non-Kramers nature of Tm$^{3+}$, according to the following CEF analysis. At high temperatures, $T$ $\gg$ ($\overline{\Delta}$ or $\overline{J_1^{zz}+J_2^{zz}}$) $\sim$ 10 K, the CEF randomness and intersite couplings can be ignored, and the single-ion CEF excitations to higher levels become dominant. Under zero applied field, the CEF Hamiltonian that is invariant under the D$_{3d}$ point group symmetry of TmMgGaO$_4$ is given by~\cite{PhysRevLett.118.107202}
\begin{multline}
\mathcal{H}_{\rm CEF}=
B_{2}^{0}O_{2}^{0}+B_{4}^{0}O_{4}^{0}+B_{4}^{3}O_{4}^{3} \\
+B_{6}^{0}O_{6}^{0}+B_{6}^{3}O_{6}^{3}+B_{6}^{6}O_{6}^{6},
\label{eqs1}
\end{multline}
where $B_{n}^{m}$ ($n,m$ are integers and $n\geq m$) are CEF parameters that will be determined experimentally, and the Stevens operators $O_{n}^{m}$ are polynomial functions of the components of the total angular momentum operators $J_z$, $J_+$, and $J_-$ ($J_{\pm}=J_x\pm iJ_y$). The eigenvalues and eigenvectors of Eq.~(\ref{eqs1}) are given by $E_{j}$ and $\mid\!E_j\rangle$ ($j$ = 1$-$13), respectively. Under an external magnetic field of $H$ along the $x$-, $y$- or $z$-direction ($z$ is along the $c$ axis, and $x$ is along the $a$ axis), the CEF Hamiltonian can be expressed as,
\begin{equation}
\mathcal{H}_{\rm CEF}^\alpha=\mathcal{H}_{\rm CEF}-\mu_0\mu_Bg_JHJ_\alpha,
\label{eqs2}
\end{equation}
with $\alpha$ = $x$, $y$, and $z$ respectively. The eigenvalues and eigenvectors of Eq.~(\ref{eqs2}) are given by $E_{j}^{\alpha}$ and $\mid\!j,\alpha\rangle$, respectively. The single-ion dc magnetic susceptibility can be calculated by,
\begin{equation}
\chi_\alpha^{\rm CEF}=\frac{\mu_Bg_JN_A\sum_{j=1}^{13} exp(-\frac{E_{j}^{\alpha}}{k_BT})\langle j,\alpha|J_\alpha|j,\alpha\rangle}{H\sum_{j=1}^{13}exp(-\frac{E_{j}^{\alpha}}{k_BT})},
\label{eqs3}
\end{equation}
and the single-ion magnetic heat capacity under 0 T can be calculated by,
\begin{equation}
C_m^{\rm CEF}=\frac{N_A}{k_{B}T^{2}}\frac{\partial^2\ln[\sum_{j=1}^{13} \exp(-\frac{E_j}{k_{B}T})]}{\partial(\frac{1}{k_{B}T})^2}.
\label{eqs4}
\end{equation}

For TmMgGaO$_4$, $\chi_x^{\rm CEF}$ = $\chi_y^{\rm CEF}$ = $\chi_{\perp}^{\rm CEF}$ and $\chi_z^{\rm CEF}$ = $\chi_{\parallel}^{\rm CEF}$ are the calculated CEF susceptibilities perpendicular and parallel to the $c$ axis, respectively. Through the combined fit to the high-$T$ magnetic susceptibilities and heat capacity measured above 90 K [see Fig.~\ref{fig1} (c) and (d) in the main text], all of the six CEF parameters (median values), $B_{n}^{m}$, can be determined experimentally (see Table~\ref{table1}). All of the thirteen eigenvalues (the relative values) and eigenvectors of Eq.~(\ref{eqs1}) are then obtained (see Table~\ref{table2}). The resulting GS $g$ tensor naturally features the strict Ising anisotropy, $g_{\perp}^{CEF}$ = 0 and $g_{\parallel}^{CEF}$ = 12.5 (see main text).

\begin{figure}[t]
\centering
\includegraphics[width=8.6cm,angle=0]{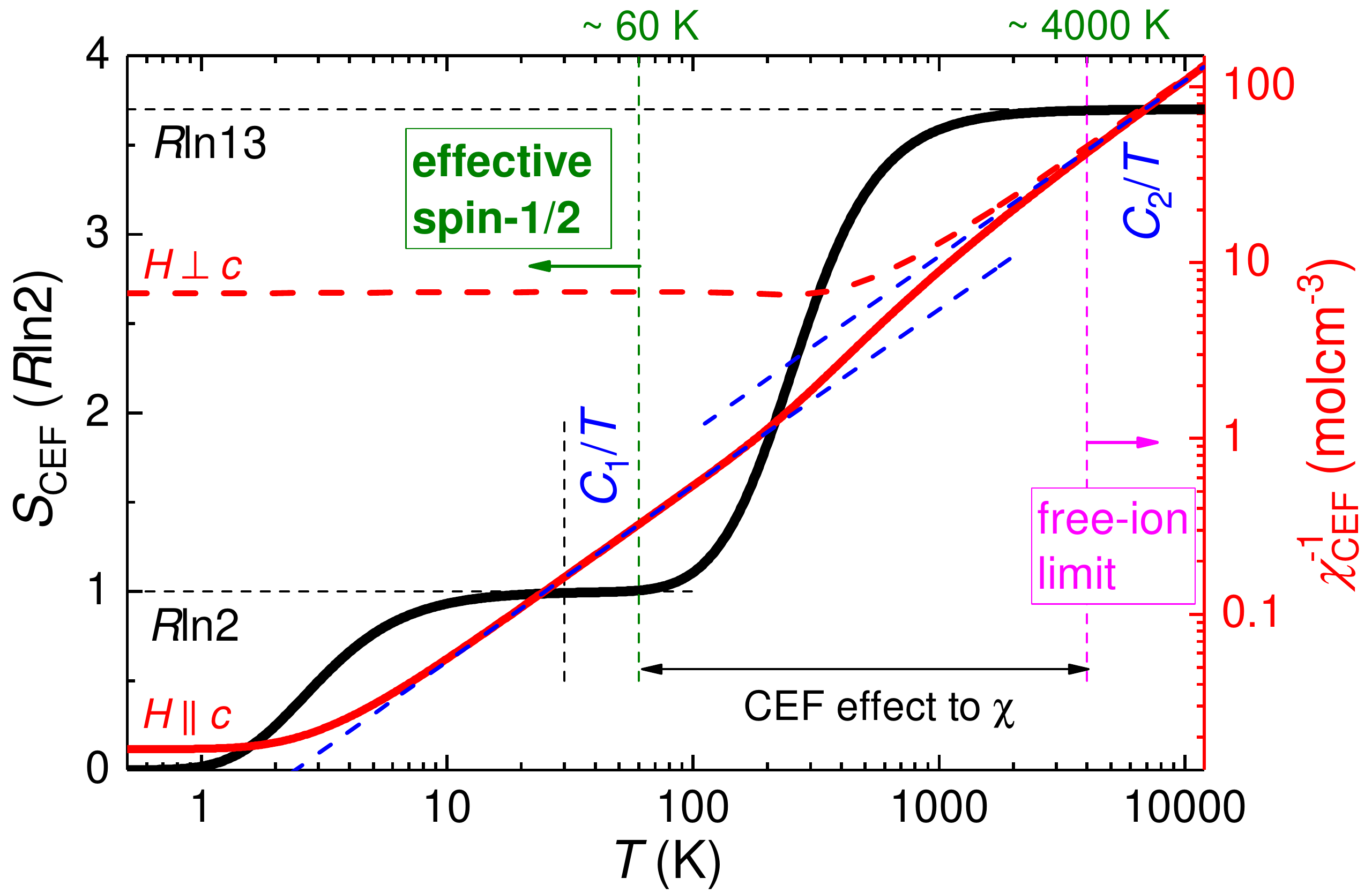}
\caption{(Color online)
Temperature dependence of the calculated CEF thermodynamic properties. The least-$R_P$ fitted average CEF parameters of TmMgGaO$_4$ are used, without any CEF randomness (broadening) and without any intersite magnetic couplings. The dashed blue lines show the Curie fits to the calculated susceptibility along the $c$ axis ($\chi_{\parallel}$ $\sim$ 1/$T$), at 30 $\leq$ $T$ $\leq$ 60 K and $T$ $>$ 4000 K, respectively. The entropy (black) and dc susceptibilities (red) are calculated at 0 and 0.05 T (measuring field), respectively.}
\label{CEF}
\end{figure}

We show the calculated CEF thermodynamic properties of TmMgGaO$_4$, without any CEF randomness (broadening) and without any intersite magnetic couplings (see Fig.~\ref{CEF}). Two robust Curie-law ($\theta_w^{\parallel}$ = 0 K) behaviors, $\chi_{\parallel}$ = $C_1$/$T$ and $C_2$/$T$, are clearly observed in two different temperature ranges at 30 $\leq$ $T$ $\leq$ 60 K and $T$ $>$ 4000 K, with the constant entropies, $S_{CEF}$ $\sim$ $R$ln2 and $R$ln13, where $C_1$ = 185 Kcm$^3$/mol $\sim$ $N_A\mu_0\mu_B^2g_{\parallel}^2$/(4$k_B$) in the effective Ising spin-1/2 range and $C_2$ = 90 Kcm$^3$/mol $\sim$ $N_A\mu_0\mu_B^2g_J^2J(J+1)$/(3$k_B$) in the high-temperature isotropic ($\chi_{\parallel}$ $\sim$ $\chi_{\perp}$) free/isolated-ion limit~\cite{Mackay1980Ferromagnetism,Dunlap1982Crystal,Dunlap1983Crystalline}, respectively (see Fig.~\ref{CEF}). In both temperature ranges, the CEF effect to the susceptibility can be neglected. Below $\sim$ 10 K, further condensation of the CEF entropy occurs due to the inner gap of the two lowest-lying singlets ($\sim$ 6.3 K).

Finally, we checked the single-ion physics by measuring specific heat of the strongly diluted sample, Tm$_{0.04}$Lu$_{0.96}$MgGaO$_4$, and observed the finite zero-temperature value, $C_m$/$T$ $\sim$ 0.65 JK$^{-2}$ per mol Tm. This indicates the mixing of the two lowest-lying CEF singlets and the formation of a quasidoublet, which renders the Ising anisotropy~\cite{nekvasil1990effective}.

\section{Neutron diffraction measurements.}
\label{a5}

\begin{figure}[t]
\centering
\includegraphics[width=8.6cm,angle=0]{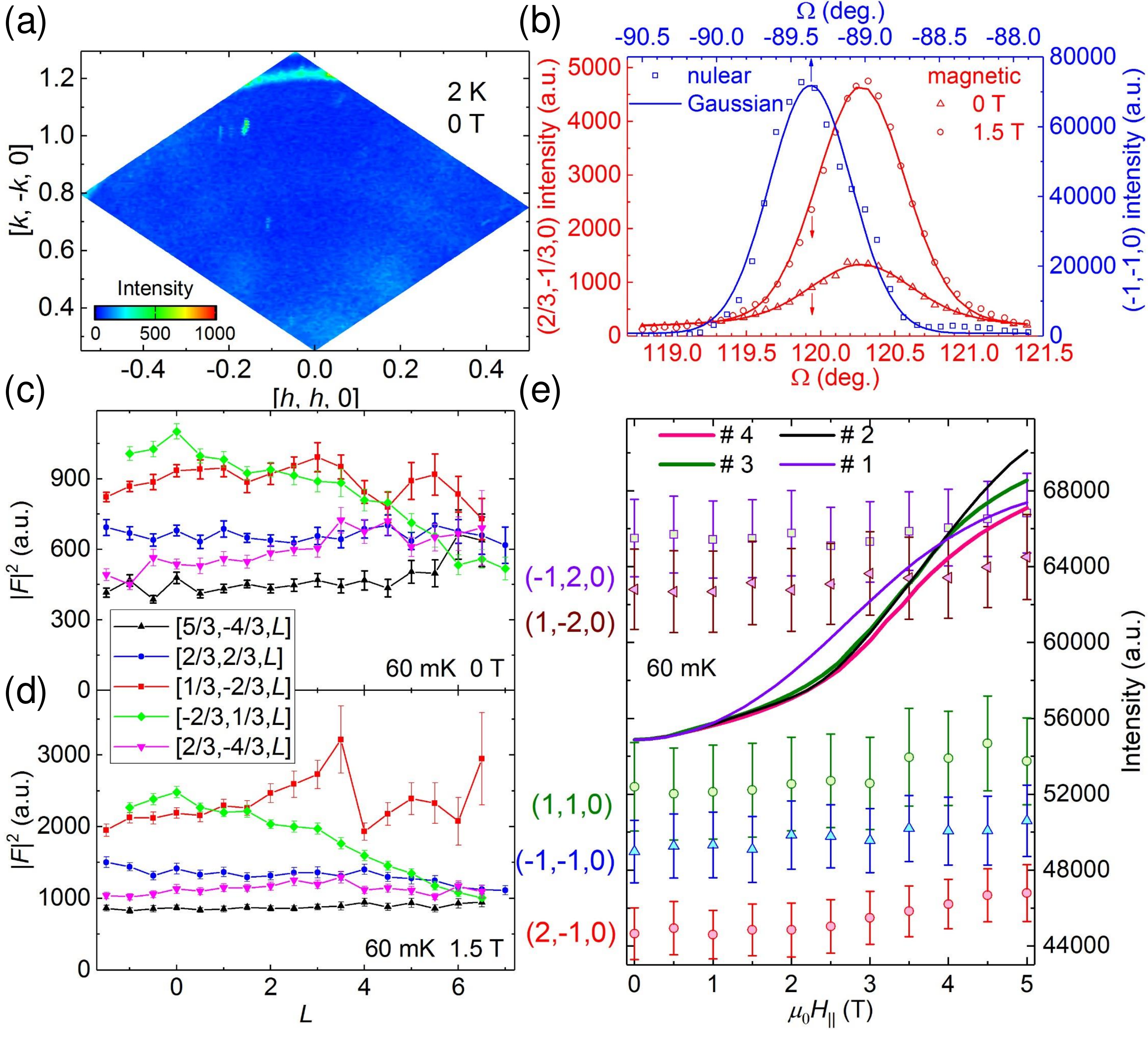}
\caption{(Color online)
(a) High-temperature neutron diffraction background measured at 2 K and 0 T on D23. The much weaker peaks are magnetic field independent at 60 mK, and shouldn't originate from the intrinsic magnetic signal of TmMgGaO$_4$. The ring-shaped signals originate from the copper sample holder. (b) $\Omega$-scans measured on the broad magnetic reflection, ($\frac{2}{3}$, -$\frac{1}{3}$, 0), in 0 and 1.5 T applied along the $c$ axis, as well as on the nuclear reflection, $(\bar 1, \bar 1, 0)$, in 0 T at 60 mK, on D23. The blue line shows the Gaussian fit with the resolution of $\sigma_{\Omega}$ = 0.65(2)$^o$, and the red lines are the fits to the data with a combination of the Gaussian and Lorentzian functions [see Eq.~(\ref{eq_n1})]. $L$ dependence of selected magnetic structure factors measured on POLI in (c) 0 T and (d) 1.5 T, at 60 mK. (e) Longitudinal magnetic field dependence of the integral intensities of five nuclear Bragg reflections measured at 60 mK on D23. The colored lines are the combinations of the field-independent nuclear contribution and the calculated magnetic part using different models. And the scale and magnetic form factors are included in the calculated magnetic contribution [see Eq.~(\ref{eqrr1})].}
\label{n1}
\end{figure}

\begin{figure}[t]
\begin{center}
\includegraphics[width=8.6cm,angle=0]{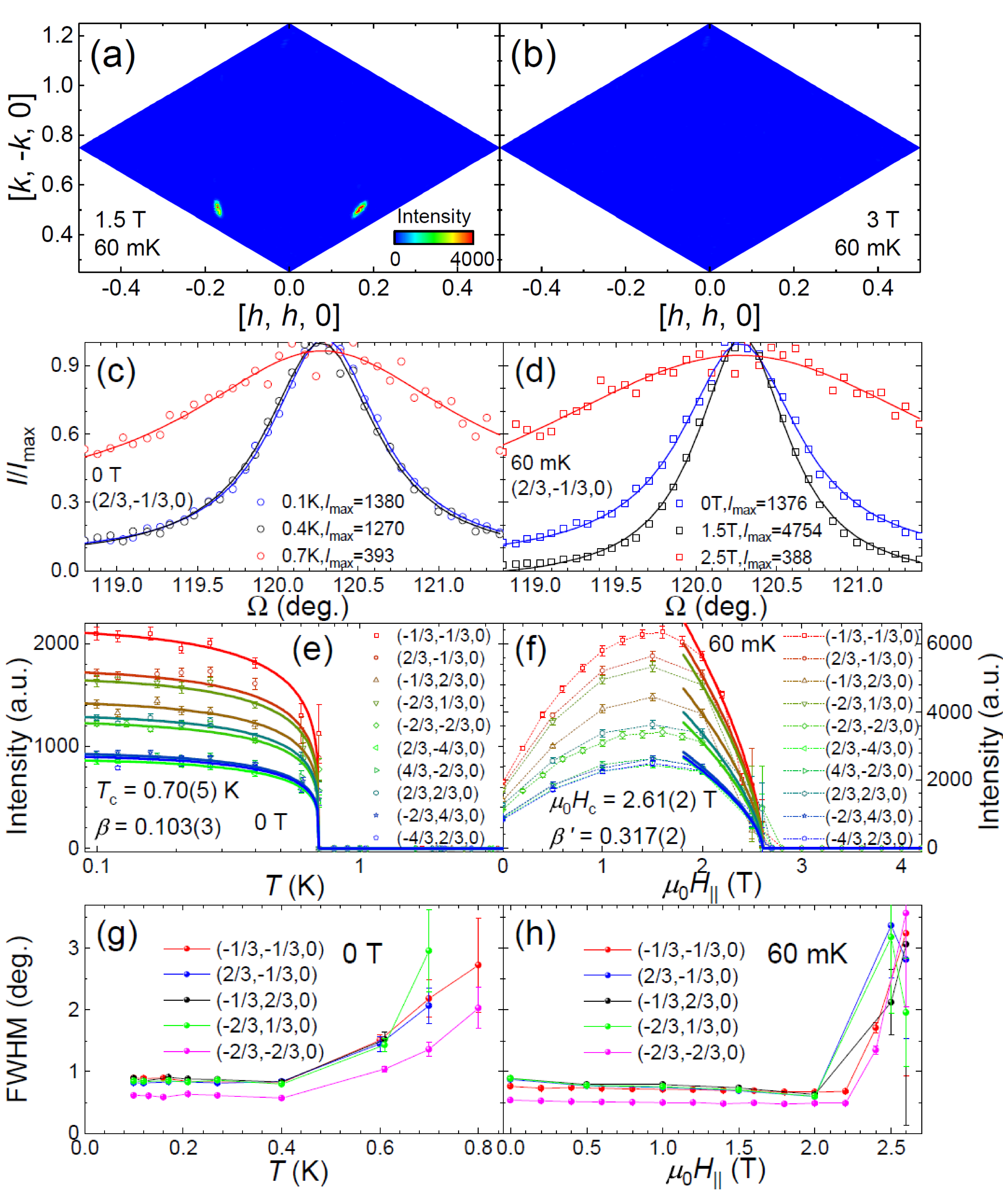}
\caption{(Color online)
Magnetic neutron diffraction of TmMgGaO4 measured on D23 at 60 mK under the magnetic field of (a) 1.5 T and (b) 3 T, applied along the $c$ axis. Selected $\Omega$-scans measured on the ($\frac{2}{3}$,$-\frac{1}{3}$,0) magnetic reflection (c) at 0 T and (d) at 60 mK. The colored lines show the corresponding Lorentzian fits. (e) Temperature and (f) magnetic field dependence of the magnetic reflection intensities measured at 0 T and at 60 mK, respectively. The colored lines show the combined critical fits. (g) Temperature and (h) magnetic field dependence of the magnetic reflection FWHMs measured at 0 T and at 60 mK, respectively.}
\label{figs7}
\end{center}
\end{figure}

\begin{table}[t]
\caption{Intensities of the magnetic reflections measured at 60 mK.}\label{table3}
\begin{center}
\begin{tabular}{ l | l | l || l | l | l }
    \hline
    \hline
   & at 0 T & at 1.5 T &  & at 0 T & at 1.5 T\\ \hline
 (-$\frac{1}{3}$,-$\frac{1}{3}$,0) & 1956(47) & 6339(168) & ($\frac{2}{3}$,-$\frac{1}{3}$,0) & 1818(54) & 5640(143) \\
 (-$\frac{1}{3}$,$\frac{2}{3}$,0) & 1405(38) & 4436(118) & (-$\frac{2}{3}$,$\frac{1}{3}$,0) & 1710(46) & 5332(151) \\
 (-$\frac{2}{3}$,-$\frac{2}{3}$,0) & 1136(27) & 3404(117) & ($\frac{2}{3}$,-$\frac{4}{3}$,0) & 868(33) & 2447(93) \\
 ($\frac{4}{3}$,-$\frac{2}{3}$,0) & 911(24) & 2632(97) & ($\frac{2}{3}$,$\frac{2}{3}$,0) & 1282(28) & 3629(119) \\
 (-$\frac{2}{3}$,$\frac{4}{3}$,0) & 917(24) & 2627(89) & (-$\frac{4}{3}$,$\frac{2}{3}$,0) & 839(26) & 2489(80) \\
 (-$\frac{1}{3}$,-$\frac{4}{3}$,0) & 675(26) & 1743(64) & ($\frac{1}{3}$,-$\frac{5}{3}$,0) & 645(37) & 1768(75) \\
 ($\frac{4}{3}$,-$\frac{5}{3}$,0) & 729(30) & 2056(86) & ($\frac{5}{3}$,-$\frac{4}{3}$,0) & 607(21) & 1776(87) \\
 ($\frac{5}{3}$, -$\frac{1}{3}$,0) & 732(32) & 1998(94) & ($\frac{4}{3}$,$\frac{1}{3}$,0) & 1088(43) & 2853(130) \\
 ($\frac{1}{3}$,$\frac{4}{3}$,0) & 699(105) & 1470(97) & (-$\frac{4}{3}$,$\frac{5}{3}$,0) & 728(38) & 2028(85) \\
 (-$\frac{5}{3}$,$\frac{4}{3}$,0) & 645(33) & 1697(63) & (-$\frac{4}{3}$,-$\frac{1}{3}$,0) & 886(40) & 2299(92) \\
 (-$\frac{2}{3}$,-$\frac{5}{3}$,0) & 351(28) & 998(79) & ($\frac{5}{3}$,$\frac{2}{3}$,0) & 1228(339) & 1365(223) \\
 ($\frac{2}{3}$,$\frac{5}{3}$,0) & 687(90) & 1458(153) & (-$\frac{5}{3}$,-$\frac{2}{3}$,0) & 155(40) & 878(107) \\
    \hline
    \hline
\end{tabular}
\end{center}
\end{table}

Neutron diffraction in the $ab$ plane ($L$ = 0) was measured on the D23 diffractometer at Institut Laue-Langevin (ILL), France, with the PG (002) monochromator ($E_i$ = 14.64 meV and $\lambda_i$ = 2.364 {\AA}) on a single crystal of TmMgGaO$_4$ (2.5$\times$5.7$\times$9.8 mm$^3$ and 0.711 g). Experiments down to 50 mK and up to 5 T applied field were performed using the dilution insert for the 12 T magnet. Neutron diffraction along the $c$ axis ($L$ $\neq$ 0) was measured on the POLI diffractometer at Heinz Maier-Leibnitz Zentrum (MLZ), Germany, with the Si (311) monochromator ($E_i$ = 62.07 meV and $\lambda_i$ = 1.148 {\AA}) on the same single crystal. Measurements down to 60 mK and up to 1.5 T were performed using the dilution insert for the 2.2 T magnet.

The nuclear Bragg reflections, $(1, 1, 0)$, $(\bar 1, \bar 1, 0)$, $(1, \bar 2, 0)$, $(\bar 1, 2, 0)$, and $(2, \bar 1, 0)$, with the integral intensity of $\sim$ 60000 and the Gaussian FWHM of $\sigma_{\Omega}$ = 0.65(2)$^o$ [instrumental resolution with PG~\cite{ressouche1999new}, see Fig.~\ref{n1} (b)], were measured on D23, and $a$ = 3.4097 {\AA} was refined below 5 K. The $(\bar 2, 1, 0)$ reflection could not be measured due to the beam shielding by the magnet.

The maps measured at 0 T covered the 0.25 $\leq H\leq$ 1.25 (0.01 per step) and -1.34 $\leq K\leq$ -0.25 (0.01 per step) range, see Fig.~\ref{n1} (a) for an example. At 1.5 and 3 T, the maps covered the 0.25 $\leq H\leq$ 1.25 (0.01 per step) and -1.25 $\leq K\leq$ -0.25 (0.01 per step) range [see Fig.~\ref{figs7} (a) and (b)]. The high-temperature background was measured on the sample at 2 K and 0 T, where the spin system is paramagnetic [see Fig.~\ref{n1} (a)], and subtracted from the low-$T$ data.

In order to evaluate the correlation length of the three-sublattice magnetic order, we choose a broad magnetic Bragg peak, ($\frac{2}{3}$, -$\frac{1}{3}$, 0) [see Fig.~\ref{figs7} (g) and (h)], measured in both 0 and 1.5 T at 60 mK. We performed the least-square fits to the data using a combination of the Gaussian and Lorentzian functions [see Fig.~\ref{n1} (b)],
\begin{equation}
 I(\Omega)=I_{bckgr}+\int Lor(\Omega')G(\Omega-\Omega')d\Omega'.
\label{eq_n1}
\end{equation}
Here, the Gaussian part $G(\Omega-\Omega')$ = $\frac{\exp[-4ln2(\Omega-\Omega')^2/\sigma_{\Omega}^2]}{\sigma_{\Omega}\sqrt{\pi/4/ln2}}$ with the fixed $\sigma_{\Omega}$ = 0.65$^o$ is due to the instrumental broadening, the Lorentzian part $Lor(\Omega')$ = $\frac{2I_0}{\pi}\frac{\omega_L}{4(\Omega'-\Omega_0)^2+\omega_L^2}$ is the intrinsic scattering signal from the three-sublattice magnetic order, and $I_{bckgr}$, $I_0$, $\omega_L$, $\Omega_0$ are fitting parameters for the background, integral intensity, intrinsic reflection width, peak center, respectively. We obtained $\omega_L$ $\sim$ 0.34$^o$ and 0.08$^o$ at 0 and 1.5 T, respectively. If we fit the magnetic reflections using a single Lorentzian function, FWHM = 0.87(2)$^o$ and 0.69(1)$^o$ are obtained at 0 and 1.5 T, respectively, with FWHM $<$ $\omega_L$+$\sigma_{\Omega}$.

Therefore, at 60 mK the correlation length of the three-sublattice magnetic order can be estimated as, $\xi_{ab}$ ({\AA}) $\sim$ 2$\pi$/[FWHM$_Q$ ({\AA}$^{-1}$)] $\sim$ $\lambda_i$/(2$\omega_L$sin$\theta$)~\cite{PhysRevB.70.214434,PhysRevB.88.024411}. With $\lambda_i$ = 2.364 {\AA} and $\theta$ = 13.4$^o$, we obtain $\xi_{ab}$ $\sim$ 850 and 3800 {\AA} at 0 and 1.5 T, respectively. The measured $\xi_{ab}$ is more than two orders of magnitude larger than the lattice parameter, $a$ = 3.4097 {\AA}, but it is still much smaller than the crystal size. Along the $c$ axis, the interlayer correlation length, $\xi_c$, can be estimated as, $\xi_c$ $\sim$ 2$\pi$/FWHM$_L$ $<$ $c$/12, where FWHM$_L$ $>$ 12$\frac{2\pi}{c}$ is the broadening of the magnetic reflections along $L$ [see Fig.~\ref{n1} (c) and (d)].

The magnetic field dependence of the intensity on the nuclear reflections measured at $\sim$ 60 mK up to 5 T is shown in Fig.~\ref{n1} (e). A rapid increase of the intensity is observed, especially on the reflections of (1,1,0) and (2,-1,0), at $\sim$ 3 T with increasing the applied field, which seems consistent with the \#2, \#3, and \#4 models [Fig.~\ref{n1} (e)]. However, the measured increase of the intensity (with large error bars) up to 5 T seems smaller than the expected values. The increase of the intensity mainly reflects the uniform/bulk magnetization process [Fig.~\ref{fig2} (c)]. The spin system of TmMgGaO$_4$ is almost fully polarized at 5 T, and thus the overall increase of the magnetic intensity with integer indexes can't be tiny [see Fig.~\ref{n1} (e)]. One possible explanation is that the dominant nuclear part may weakly depend on the applied magnetic field through the magnetostriction effect owing to the spin-lattice coupling.

At $\sim$ 60 mK, the magnetic reflections show the maximum intensities at $\mu_0H_{\parallel}$ $\sim$ 1.5 T, while completely disappear at $\mu_0H_{\parallel}$ $\sim$ 3 T. Well below the critical points, $T_c$ = 0.7 K and $\mu_0H_{c}$ = 2.6 T, the magnetic reflections are coherent with a correlation length of $\geq$ 1000 {\AA} [see Fig.~\ref{figs7} (c) and (d)].

Integral intensities of the magnetic reflections measured at 60 mK in both 0 and 1.5 T, with $|H|$ $\leq$ 2 and $|K|$ $\leq$ 2, are listed in Table~\ref{table3}. Three reflections, ($\frac{1}{3}$, $\frac{1}{3}$, 0), (-$\frac{5}{3}$, $\frac{1}{3}$, 0), and (-$\frac{1}{3}$, $\frac{5}{3}$, 0), were unavailable due to the beam shielding by the magnet, whereas the ($\frac{1}{3}$, -$\frac{2}{3}$, 0) reflection was clearly observed in the maps (see main text), but lost in $\Omega$-scans. The magnetic diffraction intensity is calculated to be $\sim$ 1400 (\#3 model) and 1100 (\#4) using Eq.~(\ref{eqs7}) and (\ref{eqrr1}), which are largely consistent with the average intensity of the magnetic reflections with $|\textbf{Q}|$ = 3.5276$\pi$/$a$ [(-1/3,-4/3,0), (1/3,-5/3,0), and so on] ($\sim$ 1900, see Table~\ref{table3}), at 1.5 T. The measured magnetic intensity decreases from $\sim$ 5400 (averaged) at $|\textbf{Q}|$ = 4$\pi$/(3$a$) to $\sim$ 1900 (averaged) at $|\textbf{Q}|$ = 3.5276$\pi$/$a$ (Table~\ref{table3}), while the magnetic form factor only slightly decreases from 0.9354 to 0.6446. Therefore, other intensity correction factors play an important role~\cite{larson1994gsas}, and we multiply the magnetic structure factor calculated with Eq.~(\ref{eqs7}) by an estimated factor of $\sim$ (0.54$\times$10$^{\text{-}12}$cm)$^2$$\times$5400$\times$0.6446/(1900$\times$0.9354)$S_{ph}$ $\sim$ 1800 Tm in Fig.~\ref{fig2} (d).

We performed the combined fit to the intensities of the magnetic reflections measured at 0.1$-$3.5 K in the field of 0 T,
\begin{equation}
 I_{HK0}=A_{HK0}\frac{|T_c-T|^{2\beta}}{1+e^{\frac{T-T_c}{T_0}}},
\label{eqs9}
\end{equation}
by sharing the same fitting parameters, $T_c$ and $\beta$. We fixed $T_0\equiv$ 0.001 K to ensure the conditional function, whereas $A_{HK0}$ were the fitted pre-factors for the reflections ($H$,$K$,0). Through the combined fit [see Fig.~\ref{figs7} (e)], the critical temperature and exponent, $T_c$ = 0.70(5) K and $\beta$ = 0.103(3), were obtained. Similarly, we also fitted the intensities measured in the fields of 2$-$5 T applied along the $c$ axis at 60 mK,
\begin{equation}
 I_{HK0}=A'_{HK0}\frac{|H_c-H|^{2\beta'}}{1+e^{\frac{H-H_{c}}{H_0}}},
\label{eqs10}
\end{equation}
by sharing the same fitting parameters, $H_{c}$ and $\beta'$. We fixed $\mu_0H_0\equiv$ 0.001 T, and obtained $\mu_0H_{c}$ = 2.61(2) T and $\beta'$ = 0.317(2) [see Fig.~\ref{figs7} (f)].

Around (just below) the critical points, the (quasi-)long-range spin order is replaced by the short-range one, and thus the FWHM of the magnetic reflections increases quickly [see Fig.~\ref{figs7} (g) and (h)]~\cite{zaliznyak2004magnetic}.

\begin{figure}[t]
\begin{center}
\includegraphics[width=8.6cm,angle=0]{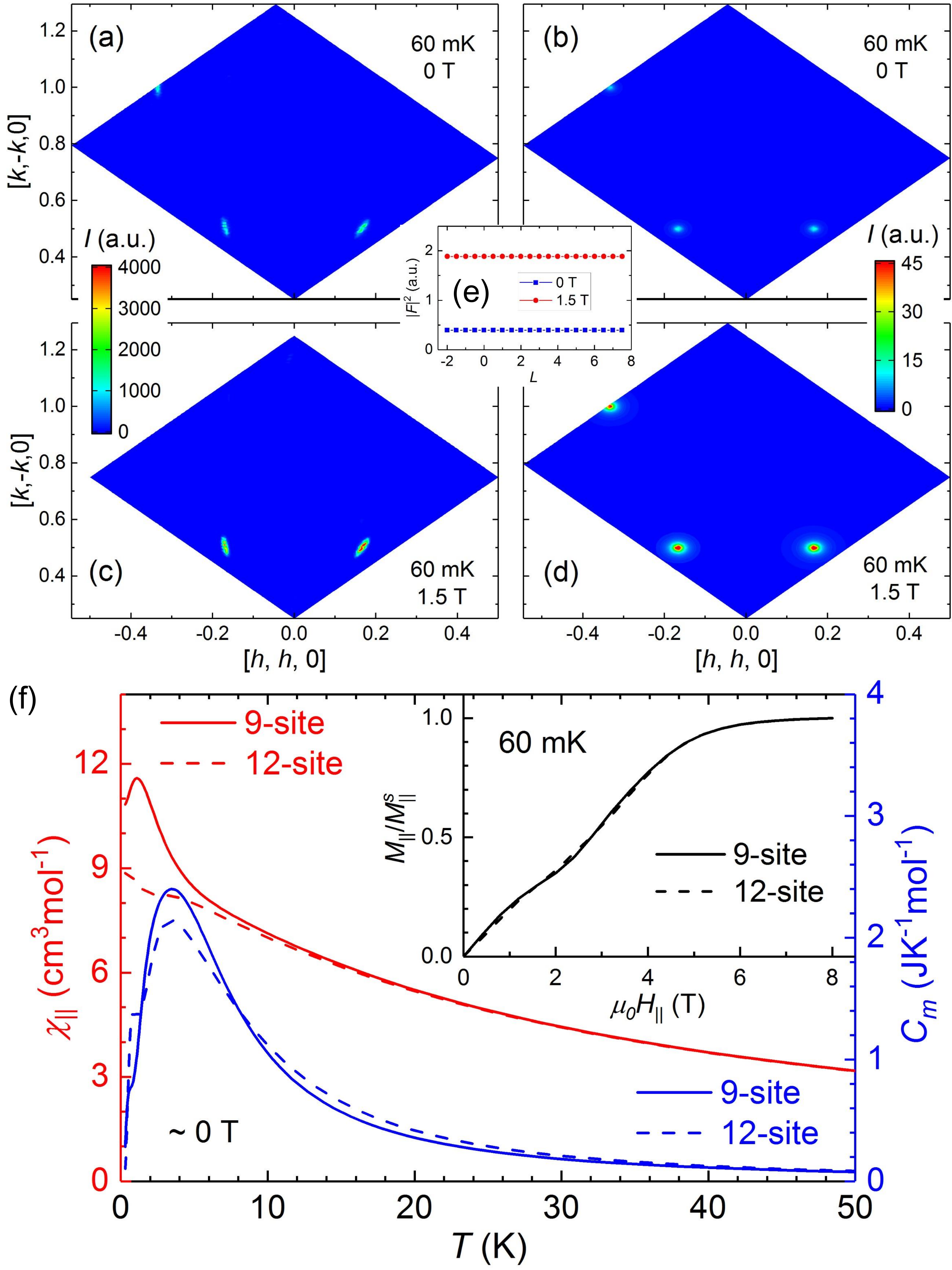}
\caption{(Color online)
Magnetic neutron diffraction of TmMgGaO$_4$ measured at 60 mK in (a) 0 T and (c) 1.5 T. The calculated spectra are shown at (b) 0 T and (d) 1.5 T using model \#4 by Eq.~(\ref{eqs7}) with the magnetic form factor. The isotropic resolution is used in the momentum (\textbf{Q}) space, $\sim$ 0.015, in both (b) and (d). (e) Static structure factor per Tm calculated by model \#4 [see Eq.~(\ref{eqs7})] along [$\frac{1}{3}$, -$\frac{2}{3}$, $L$] at 0 and 1.5 T, at 60 mK. (f) Thermodynamic properties calculated on the 9-site (solid lines) and 12-site (dashed lines) clusters with different PBC. The same Hamiltonian parameters of the fitted model \#2 are used. The red, blue, and black (the inset) lines show the calculated dc susceptibility at 0.1 T (measuring field), heat capacity at 0 T, and magnetization at 60 mK, respectively.}
\label{n2}
\end{center}
\end{figure}

\section{Exact diagonalization calculations and simulations.}
\label{a4}

\begin{figure}[t]
\begin{center}
\includegraphics[width=8.6cm,angle=0]{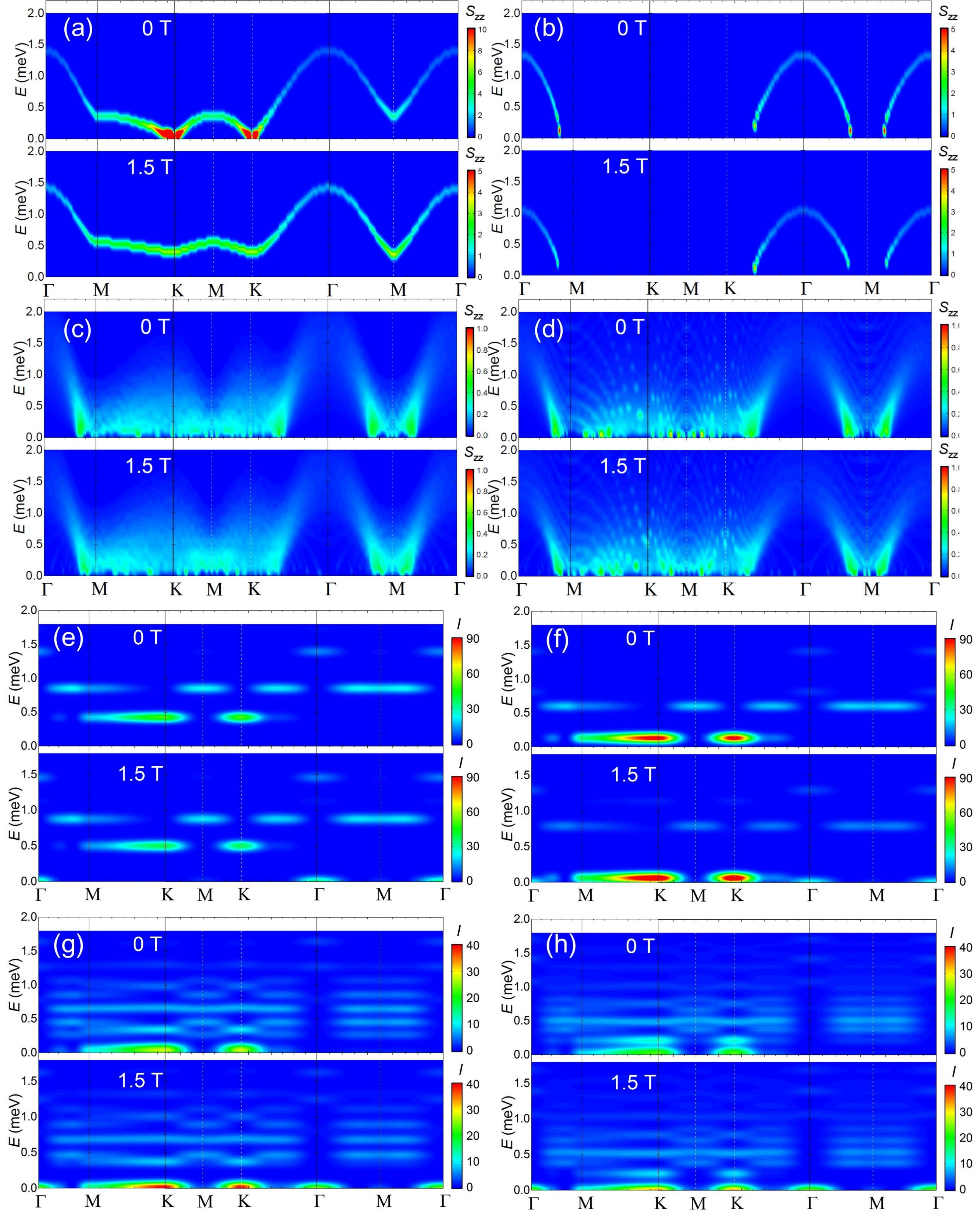}
\caption{(Color online)
Calculated spin-wave excitations by Spinw at the longitudinal fields of 0 T and 1.5 T, using (a) \#1, (b) \#2, (c) \#3, and (d) \#4 models. Calculated INS spectra by the ED using the 9-site cluster with PBC at the longitudinal fields of 0 T and 1.5 T, at 50 mK, using (e) \#1, (f) \#2, (g) \#3, and (h) \#4 models [see Eq.~(\ref{eqs8})]. The instrumental resolution of 0.114 meV is used at $E_i$ = 4.8 meV, and the linear \textbf{Q}-scan ranges are kept the same as Ref.~\cite{shen2018hidden}.}
\label{sw}
\end{center}
\end{figure}

Including the precise single-ion and bond disorder effects into the many-body correlated model of TmMgGaO$_4$ is a challenging problem. For simplicity, we kept all symmetries of the system (space group: $R\overline{3}m$), and assumed the distributions $Y$-$\overline{Y}$ = $K_Y$($\Delta$-$\overline{\Delta}$) around the average value in both \#3 and \#4 models (see main text), where $Y$ = $J_1^{zz}$, $J_2^{zz}$, $g_{\parallel}$ is the Hamiltonian parameter, and $K_Y$ is the fitting linear parameter proportional to the FWHM of $Y$. Each set of the Hamiltonian parameters corresponds to one local CEF environment (Mg$^{2+}$/Ga$^{3+}$ arrangement), and we perform the ED calculations for each of these sets. To facilitate the calculations, we truncated the distribution function at $|\Delta$-$\overline{\Delta}|$/FWHM($\Delta$) $\geq$ 1.3 and 2.0 with $P$($\Delta$)/$P$($\overline{\Delta}$) $\leq$ 0.9\% and $\leq$ 5.9\%, for \#3 (Gaussian) and \#4 (Lorentzian) models, respectively, and then normalized the numerical distribution function by $\sum_{\gamma}P$($\Delta_{\gamma}$) = 1.

For each set of the Hamiltonian parameters, we calculate the magnetization [see Fig.~\ref{fig2} (c) in main text] using
\begin{multline}
M_{\parallel}= \\
\frac{\mu_Bg_{\parallel}\sum_jexp(-\frac{E_j(H_{\parallel})}{k_BT})\langle j,H_{\parallel}| \sum_nS_n^z | j,H_{\parallel}\rangle}{N\sum_jexp(-\frac{E_j(H_{\parallel})}{k_B T})},
\label{eqs5}
\end{multline}
where $E_j(H_{\parallel})$ and $|j,H_{\parallel}\rangle$ are the eigenvalue and eigenstate of Eq.~(\ref{Eq5}) (see main text), after the ED calculation. The dc magnetic susceptibility is obtained as $\chi_{\parallel}$ = $N_AM_{\parallel}$/$H_{\parallel}$ [see Fig.~\ref{fig2} (a) in main text]. The zero-field heat capacity [see Fig.~\ref{fig2} (b) in main text] can be calculated as
\begin{equation}
C_m=\frac{N_A}{Nk_{B}T^{2}}\frac{\partial^2\ln[\sum_j \exp(-\frac{E_j(H_{\parallel}=0)}{k_{B}T})]}{\partial(\frac{1}{k_{B}T})^2}.
\label{eqs6}
\end{equation}
And the static structure factor of the Ising dipole moment [see Fig.~\ref{fig2} (d) in main text] is calculated by
\begin{multline}
|F|^2 \sim  \\
\frac{\sum_jexp(-\frac{E_j(H_{\parallel})}{k_BT})|\langle j,H_{\parallel}| \sum_ng_{\parallel}S_n^zexp(i\mathbf{Q}\cdot\mathbf{r}_n) |j,H_{\parallel}\rangle|^2}{N^2\sum_jexp(-\frac{E_j(H_{\parallel})}{k_B T})},
\label{eqs7}
\end{multline}
where $\mathbf{r}_n$ is the position vector of the $n$th site on the triangular lattice. Therefore, the magnetic neutron diffraction intensity can be further calculated as~\cite{larson1994gsas}
\begin{equation}
I=(0.54\times10^{\text{-}12}\text{cm})^2S_{ph}|f(|\mathbf{Q}|)|^2|F|^2,
\label{eqrr1}
\end{equation}
where $S_{ph}$ $\sim$ ($I_{(\text{-}1,2,0)}$+$I_{(1,\text{-}2,0)}$+$I_{(1,1,0)}$+$I_{(\text{-}1,\text{-}1,0)}$ +$I_{(2,\text{-}1,0)}$)/5/$|F_n|^2$ $\sim$ 3.0$\times$10$^{27}$ Tm cm$^{\text{-}2}$ is the scale factor at $|\textbf{Q}|$ = 4$\pi$/$a$ obtained from the nuclear reflections measured at 0 T and 60 mK [see Fig.~\ref{n1} (e)], and $|F_n|^2$ = 1.84$\times$10$^{\text{-}23}$ cm$^2$/Tm is the structure factor of these reflections calculated with the reported crystal structure of TmMgGaO$_4$~\cite{cevallos2017anisotropic}. And $|f(|\mathbf{Q}|)|^2$ is the magnetic form factor of Tm$^{3+}$.

Finally, the observables, $X_i^{cal}$, is obtained by, $X_i^{cal}$ = $\sum_{\gamma}P$($\Delta_{\gamma}$)$X_i$($\Delta_{\gamma}$), in \#3 and \#4 models. Using Eq.~(\ref{eqs7}), we can largely reproduce the low-$T$ magnetic neutron diffraction measured on the single crystal of TmMgGaO$_4$ [see Fig.~\ref{n2} (a) - (e)].

We also perform the ED calculation on the 12-site cluster with different PBC [see Fig.~\ref{fig4} (b) for the geometry]. The calculated thermodynamic data are shown in Fig.~\ref{n2} (f), with the previous 9-site ED result for comparison. Although certain differences are observed at low temperatures, the overall trend is similar. The measured signal to noise ratio (the standard deviation) of the magnetic heat capacity is much larger than that of the magnetization (susceptibility) data due to the technical difference (Fig.~\ref{fig2}), and thus the slight difference [Fig.~\ref{n2} (f)] in the magnetic heat capacity obtained on different clusters won't significantly affect the final (fitted) result [please see Eq.~(\ref{Eq1})]. Indeed, our best parameterization ($\Delta$ = 5.7 K, $J_1^{zz}$ = 10.9 K, $g_{\parallel}$ = 13.6, $J_2^{zz}$ = 1.1 K) shows excellent agreement with the theoretical result reported in the recent preprint (after the initial submission of our present work), where the authors used quantum Monte Carlo (QMC) method and arrived at $\Delta$ = 0.54$J_1^{zz}$ = 6.2 K, $J_1^{zz}$ = 0.99 meV = 11 K, $g_{\parallel}$ = 1.101$\times$12 = 13.2, $J_2^{zz}$ = 0.05$J_1^{zz}$ = 0.6 K~\cite{li2019ghost}, using the same model (\#2). At relatively high temperatures and/or in high longitudinal magnetic fields, the size effect is relatively small. On the other hand, at low temperatures ($\leq$ 1 K) and at $\sim$ 0 T our ED calculation becomes semi-quantitative. Therefore, we get a relatively large deviation from the QMC result on $J_2^{zz}$~\cite{li2019ghost}, because this coupling mostly affects the low-energy part of the spectrum. Therefore, we mainly focus on the low-$T$ ($\sim$ 60 mK) physics of TmMgGaO$_4$ in magnetic fields $\sim$ 1.5 T ($\mu_0H_{\parallel}g_{\parallel}\mu_B$/$k_B$ $\sim$ 13 K), where the up-up-down order is most stable and the ED calculations should be accurate enough.

The spin-wave excitations can be calculated by the Spinw-Matlab code based on the linear spin-wave theory~\cite{toth2015linear} [see Fig.~\ref{sw} (a) - (d)]. With the above code, the \#1 model largely reproduces the spin-wave excitation measured at 0 T, which is sensitive to the main nonmagnetic phase with large inner gaps, while it completely fails to explain the thermodynamic properties and magnetic neutron diffraction measured under the longitudinal field (see Fig.~\ref{fig2} in main text). For example, this calculated spin-wave excitations of the \#1 model show a full gap of $\geq$ 0.4 meV, and obviously can't account for the highly enhanced reflections at K points anymore, at $\sim$ 1.5 T [see Fig.~\ref{sw} (a)]. Moreover, the measured width of the spin-wave excitation seems much wider than the reported instrumental resolution ($\sigma_E$ = 0.114 meV) at $E_i$ = 4.8 meV~\cite{shen2018hidden}.

Very recently, Ref.~\cite{li2019ghost} pointed out that the linear spin-wave approximation may get invalid in TmMgGaO$_4$. Similarly, we also calculate the INS spectra using the ED results as
\begin{multline}
I(\mathbf{Q},E) \sim \frac{g_{\parallel}^2|f(|\mathbf{Q}|)|^2k_f}{N^2k_i}\sum_{j,j',n,n'}\frac{exp(-\frac{E_j(H_{\parallel})}{k_BT})}{Z(H_{\parallel},T)} \\
\times\langle j,H_{\parallel}|S_{n}^ze^{-i\mathbf{Q}\cdot\mathbf{r}_n} |j',H_{\parallel}\rangle\langle j',H_{\parallel}|S_{n'}^ze^{i\mathbf{Q}\cdot\mathbf{r}_{n'}} |j,H_{\parallel}\rangle \\
\times\frac{\exp(\frac{-4ln2(E+E_j(H_{\parallel})-E_{j'}(H_{\parallel}))^2}{\sigma_E^2})}{\sigma_E\sqrt{\frac{\pi}{4ln2}}}.
\label{eqs8}
\end{multline}
Here, $k_i$ and $k_f$ are the incident and final neutron wave-vectors, and $Z(H_{\parallel},T)$ is the partition function. By setting $E$ = 0 meV ($k_i$ = $k_f$) and $\sigma_E$ = 0 meV, Eq.~(\ref{eqs8}) is equivalent to Eq.~(\ref{eqs7}) (the integral structure factor of the $\Omega$-scan of the neutron diffraction), after the normalization by the magnetic form factor. Due to the size effect of the ED calculation, the resulted resolution of the transfer momentum is very low [see Fig.~\ref{sw} (e) - (h)]. While, at K points our calculated diffraction intensities using \#1 and \#2 models and at 0 T by the ED are well consistent with the recently reported QMC results equipped with stochastic analytical continuation~\cite{li2019ghost} [see Fig.~\ref{sw} (e) and (f), respectively]. Similarly, our ED calculation using the \#1 model at K points clearly contradicts with the measured spin-wave excitations~\cite{shen2018hidden}, as well as the Spinw calculation based on the linear spin-wave approximation [Fig.~\ref{sw} (a)]. As the \#1 model gives an energy gap of $\sim$ 0.42 meV at K points similar to that reported in Ref.~\cite{li2019ghost}, and much larger than $\sigma_E$ at 0 T. On the other hand, our ED calculations using the random \#3 and \#4 models well reproduce the observed (quasi-)gapless (gap $<$ $\sigma_E$) feature at K points in 0 T~\cite{shen2018hidden}, as well as the enhanced magnetic reflection intensity at $\sim$ 1.5 T [see Fig.~\ref{sw} (g) and (h)]. Therefore, we emphasize the important ingredient, the distribution of the effective spin-1/2 Hamilitonian parameters, on the correlated magnetism of non-Kramers GS quasidoublets. It is caused by the nonmagnetic Mg/Ga site-mixing disorder, which is expected to be uniformly distributed at the Mg/Ga sites in RMgGaO$_4$ (R = rare-earth), according to the diffraction and structure refinements~\cite{li2015gapless,li2015rare,cevallos2017anisotropic}. Interestingly, the coherent magnetic reflections and (quasi-)long-range magnetic order, instead of the short-range spin-glass GS, are observed in TmMgGaO$_4$, despite the above randomness. Ref.~\cite{bradley2019robust} also reported that the long-range order can survive in a triangular Ising antiferromagnet, in the presence of the uniform bond randomness.

\bibliography{Tm_1}

\end{document}